\documentclass[letter]{aa} 

\usepackage{graphicx}
\usepackage{txfonts}
\usepackage[colorlinks=true,
            linkcolor=blue,
            citecolor=blue,
            urlcolor=blue,
            pdfborder={0 0 0}]{hyperref}
\usepackage{multirow} 

\mathchardef\mhyphen="2D 

\begin{document} 

   \title{Evidence for magnetic activity at starbirth: a powerful X-ray flare from the Class~0 protostar HOPS 383}

   \titlerunning{a powerful X-ray flare from the Class~0 protostar HOPS 383}

   \author{Nicolas Grosso\inst{1}
          \and 
          Kenji Hamaguchi\inst{2,3}
          \and
          David A.\ Principe\inst{4} 
          \and Joel H.\ Kastner\inst{5}
          }
          
   \institute{Aix-Marseille Univ, CNRS, CNES, LAM, Marseille, France\\
              \email{nicolas.grosso@lam.fr}
             \and
             CRESST II and X-ray Astrophysics Laboratory NASA/GSFC, Greenbelt, MD, USA
             \and
             Department of Physics, University of Maryland, Baltimore County, Baltimore, MD, USA
             \and
             Massachusetts Institute of Technology, Kavli Institute for Astrophysics and Space Research, Cambridge, MA, USA
             \and
             Center for Imaging Science, School of Physics \& Astronomy, 
             and Laboratory for Multiwavelength Astrophysics,\\ Rochester Institute of Technology, 
             Rochester, NY, USA
             }
             
   \date{Received 16 April 2020 / Accepted 2 May 2020}

 
  \abstract
   {Class~0 protostars represent the earliest evolutionary stage of solar-type stars, 
    during which the majority of the system mass resides in 
    an infalling envelope of gas and dust and is not yet in the 
    central, nascent star.
    Although X-rays are a key signature of magnetic activity in
    more evolved protostars and young stars,
    whether such magnetic activity is present at the Class~0 stage is still debated.
    }
   {We aim to detect a bona fide Class~0 protostar in X-rays.}
   {We observed HOPS~383 in 2017~December in X-rays with 
   the \emph{Chandra} X-ray Observatory ($\sim$84~ks) 
   and in near-infrared imaging 
   with the Southern Astrophysical Research telescope.}
    {HOPS~383 was detected in X-rays during a powerful flare.
    This hard ($E>2$~keV) 
    X-ray counterpart was spatially coincident 
    with the northwest 4~cm component of HOPS~383, 
    which would be the base of the radio thermal jet 
    launched by HOPS~383.
    The flare duration was $\sim$3.3\,h; at the peak, 
    the X-ray luminosity reached 
    $\sim$$4\times10^{31}$~erg~s$^{-1}$ 
    in the 2--8~keV energy band,
    a level at least an 
    order of magnitude larger than that of the undetected quiescent emission from HOPS~383.
    The X-ray flare spectrum is highly absorbed
    ($N_\mathrm{H}\sim7\times10^{23}$~cm$^{-2}$),
    and it displays a 6.4~keV emission line 
    with an equivalent width of 
    $\sim$$1.1$~keV,
    arising from neutral or low-ionization iron.}
   {
   The detection of a powerful X-ray flare from HOPS~383 constitutes
   direct proof that magnetic activity can be present at the earliest 
   formative stages of solar-type stars.
   }


   \maketitle
%

\section{Introduction}

Low-mass objects that have evolved beyond the Class~0 stage are conspicuous X-ray emitters
\citep{dunham14}, that is, Class~I 
protostars with remnant envelopes and massive accretion 
disks, and 
Class~II and III pre-main sequence stars 
with and without circumstellar disks (T~Tauri stars).
Their high luminosities ($\sim$$10^{28-31}$\,erg\,s$^{-1}$) 
compared to the solar maximum ($\sim$$10^{27}$\,erg\,s$^{-1}$)
and their intense flaring activity in X-rays
make them appear as extremely magnetically
active young suns \citep{feigelson99,guedel09}.
In Class~0 protostars 
\citep{andre93,andre00}, the hydrostatic core is deeply embedded within 
its envelope and the molecular cloud,
making its detection difficult at most wavelengths 
\citep{giardino07b}.
The most deeply-embedded X-ray sources reported 
in star-forming regions 
(\citealt{hamaguchi05}; 
\citealt{getman07} and references therein;
\citealt{kamezaki14})
have evolved beyond the Class~0 stage 
or their bolometric luminosity is not accurate enough for 
a robust conclusion (Appendix~\ref{appendix:stage}).
Moreover, any non-thermal radio emission from 
putative magnetic activity at Class~0 protostars 
is very likely
absorbed by the bases of their ionized outflows \citep{guedel02,dzib13}.

The object of this study in the Orion Molecular Cloud~3 (OMC-3) 
was identified as a protostar by a \emph{Spitzer} survey \citep{megeath12}
and included in the Herschel Orion protostar survey \citep[HOPS;][]{furlan16}
as source 383.
The submillimeter to bolometric luminosity ratio of \object{HOPS~383}, 1.4\%, \citep{safron15} 
well surpasses the (0.5\%) threshold for bona fide Class~0 
designation \citep{andre93,andre00}.
Notably, HOPS~383 is the first Class~0 protostar known to have 
undergone a mass-accretion-driven eruption \citep{safron15}, 
which peaked by 2008 and ended by 2017 September
(Appendix~\ref{appendix:wise}).

\begin{figure*}[!t]
\centering
\renewcommand\arraystretch{0} 
\begin{tabular}{@{}l@{}ll@{}l@{}}
\raisebox{43mm}{\bf a}  & \multirow{2}{*}[4.55cm]{\includegraphics[width=98.mm]{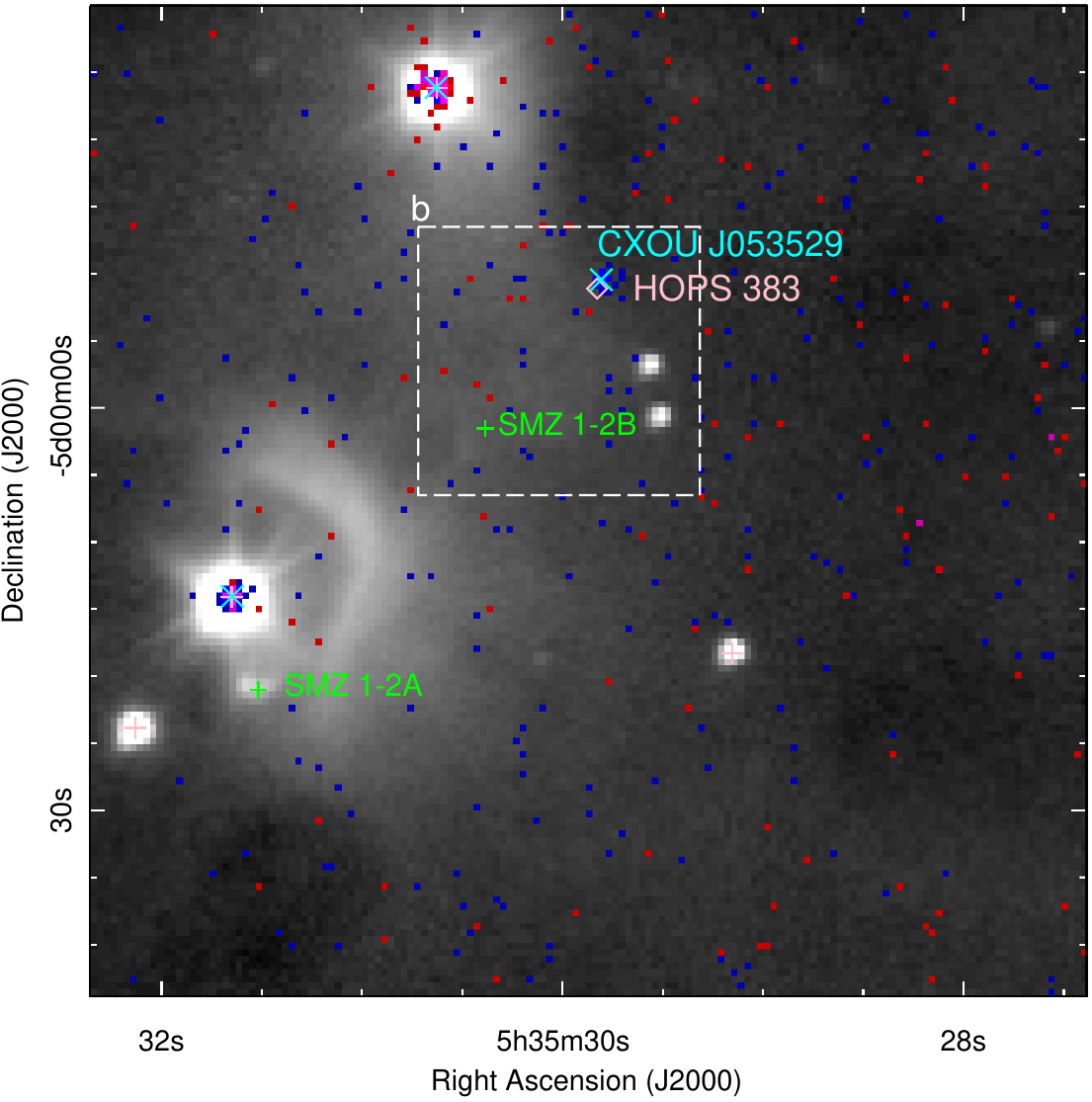}} & \raisebox{42mm}{\bf b} &\includegraphics[width=46mm]{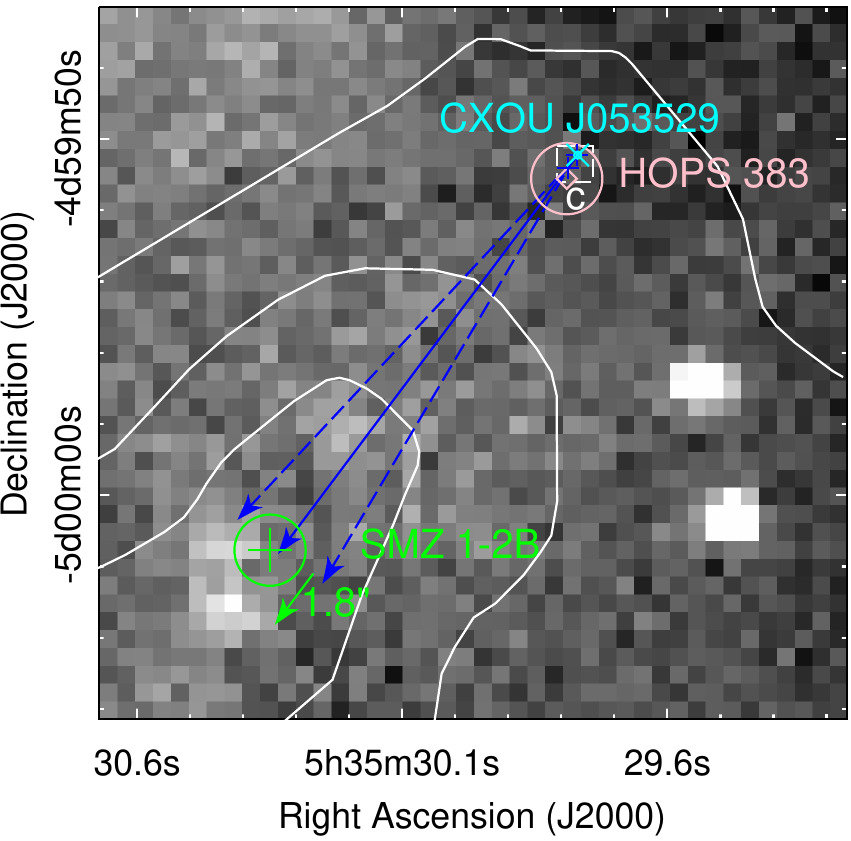}
\vspace{1mm}\\
                        &                                                                                                    & \raisebox{45mm}{\bf c} &\includegraphics[width=46mm]{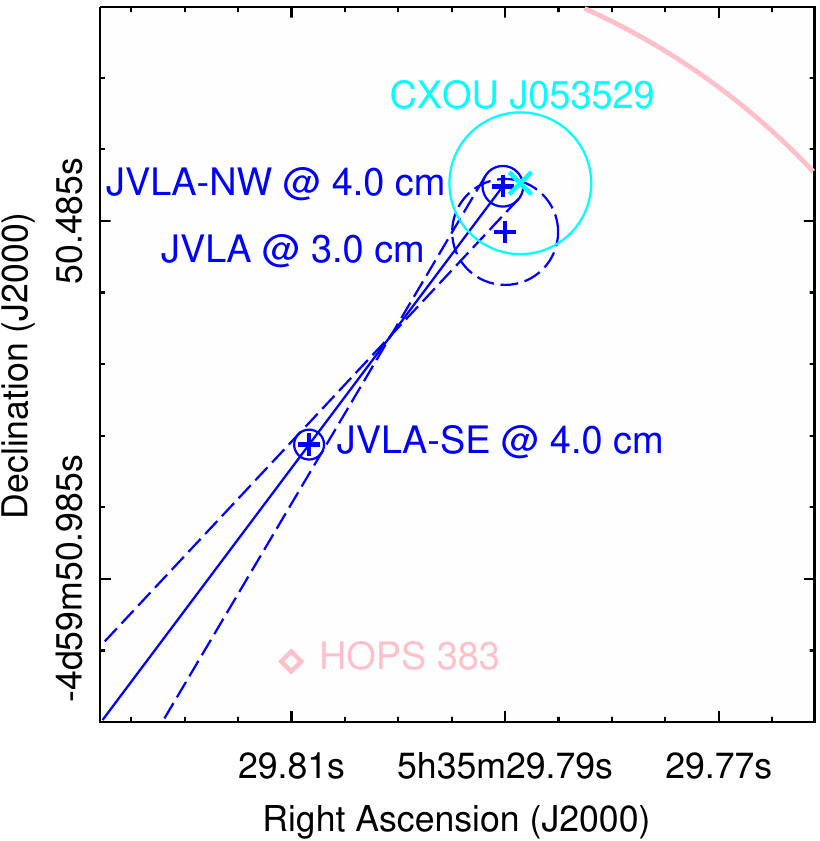}\\
\end{tabular}
\vspace{0.4cm}
\caption{\emph{Chandra} detection of HOPS~383. 
\emph{Panel~a}: 
the red and blue $0\farcs492\times0\farcs492$ pixels display logarithmically the 0.5--2~keV and 2--7~keV 
X-rays, respectively,
which were detected from 
2017~December~13 to 14
during three ACIS-I exposures, totaling 83\,877~s.
The logarithmic grayscale $K$-band image was obtained on 
2017~December~14
with the Spartan infrared camera on the SOAR telescope. 
North is up and east is left.
The pink pluses and diamond, the green pluses, and cyan crosses are 
near- and mid-infrared sources \citep{megeath12},
H$_2$-shocked emission \citep{stanke02}, and X-ray 
sources, respectively.
The white dashed box shows the 
field-of-view of Fig.~\ref{fig:fov}b.
\emph{Panel~b}: Spartan H$_2$-filter image obtained on 
2017~December~14
in linear grayscale. 
The white contours show the $^{12}$CO 
blueshifted outflow of HOPS~383 \citep{feddersen20}.
The blue pluses are radio sources \citep{galvan-madrid15,rodriguez17b}. 
The blue and
green arrows show 
the direction of the thermal-jet candidate and
the proper motion 
of the H$_2$ emission knot SMZ~1-2B, 
respectively.
The white dashed box shows the 
field-of-view
of Fig.~\ref{fig:fov}c.
\emph{Panel~c}: enlargement around HOPS~383 shows the 
mid-infrared (pink), 
radio (blue),
and X-ray (cyan)
positional error 
circles.
}
\label{fig:fov}
\end{figure*}

\section{Observations}

We observed HOPS~383 three times with the \emph{Chandra} X-ray Observatory
\citep{weisskopf02} 
 from 2017~December~13 to 14 with simultaneous
near-infrared imaging on 2017~December~14
using the 4.1~m Southern Astrophysical Research (SOAR) telescope 
\citep{krabbendam04} in Chile.

We used 
\emph{Chandra}'s
Advanced CCD Imaging Spectrometer (ACIS-I) 
in the very faint mode with a
frame time of 3.141~s (Appendix~\ref{appendix:cxo_log}).
\emph{Chandra} ACIS-I has on-axis a 
full-width half-maximum (FWHM)
angular resolution of $0\farcs5$
and a FWHM spectral resolution of 261~eV at 6.4~keV.
Data reduction is described
in Appendix~\ref{appendix:cxo_reduction}.

The SOAR Spartan infrared camera \citep{loh12} 
is composed of 4 CCDs 
of $2,048\times2,048$~pixels. 
We selected the wide-field mode with a single-detector field-of-view of $2\farcm25\times2\farcm25$
(i.e., $5\farcm04\times5\farcm04$ edge-to-edge) 
and a pixel scale of $0\farcs0661$ 
with the $K$-band and H$_2$ narrow-band filters. 
We used a five-point dithering pattern where, in the first exposure, HOPS~383 was put near the center of the 
northeastern detector (det3), 
which was then moved by $30\arcsec$ from its initial position sequentially toward the south, north, west, and east in the four other exposures.
This sequence was repeated until the requested exposure was achieved 
(Appendix~\ref{appendix:soar_log}).
Data reduction is described
in Appendix~\ref{appendix:soar_reduction}.

\section{Near-infrared results}

The near-infrared nebulosity
that was prominent during the outburst \citep{safron15}
is not detected in our SOAR $K$-band image
(Fig.~\ref{fig:fov}a), 
indeed, it vanished by 2015 December~30 \citep{fischer17c}.
The H$_2$-narrow filter image shows an emission knot of
shocked molecular hydrogen at $\sim$15\arcsec\ 
southeast from the mid-infrared location of HOPS~383.
We identify this H$_2$ source with
the H$_2$ emission knot \object{SMZ~1-2B} \citep{stanke02} that was observed on 1997~September~13,
yielding a proper motion of $\sim$$1\farcs8\pm1\arcsec$ southeast
in 20.25~yr (Fig.~\ref{fig:fov}b).
The corresponding velocity is
$\sim$$(95\pm53)\times(\sin{i}/\!\sin{69.5^\circ})^{-1}\times(d/420~\mathrm{pc})$~km~s$^{-1}$,
where $d$ is the distance to HOPS~383 and 
$i$ is the inclination angle 
\citep[$90^\circ$ corresponds to an edge-on view; 
Table~1 of][]{furlan16}. 
This is typical of 
the velocity values observed in the outflows from Class~0 protostars \citep{bontemps96,reipurth01}.
Assuming a constant velocity, the angular distance of $\sim$16\arcsec\ 
between HOPS~383 and this emission knot corresponds 
to a kinematical timescale of $\sim$$180\pm100$~yr.
The proper motion direction 
is consistent with 
the orientation of the
outburst nebula,
 the $^{12}$CO 
blueshifted outflow \citep{feddersen20},
and the position angle 
of the binary radio counterpart of HOPS~383 at 4~cm \citep{galvan-madrid15}.
We note that the unresolved 3~cm
counterpart  \citep{rodriguez17b} 
corresponds to the faint northwest component at 4~cm, \object{JVLA-NW}, 
which suggests that JVLA-NW would be the base of the radio thermal jet launched by HOPS~383,  
whereas the brighter southeast component at 4~cm, \object{JVLA-SE}, would be an emission knot along 
this thermal jet (Fig.~\ref{fig:fov}c).

\begin{figure*}[!t]
\centering
\renewcommand\arraystretch{0} 
\begin{tabular}{@{}l@{}ll@{}l@{}}
\raisebox{36mm}{\bf a}  & \multirow{2}{*}[3.9cm]{\includegraphics[width=75.mm]{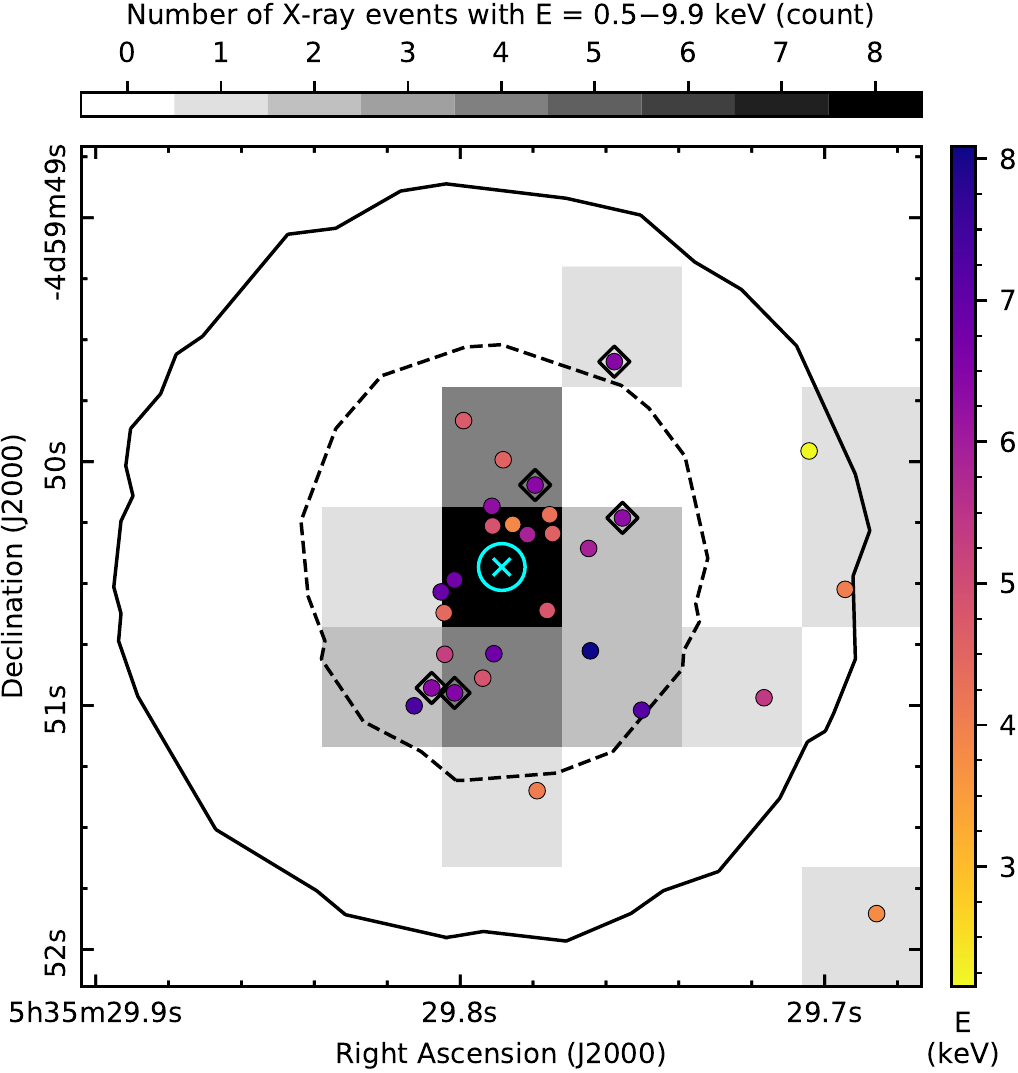}} & \raisebox{35mm}{\bf b} &\includegraphics[width=80mm]{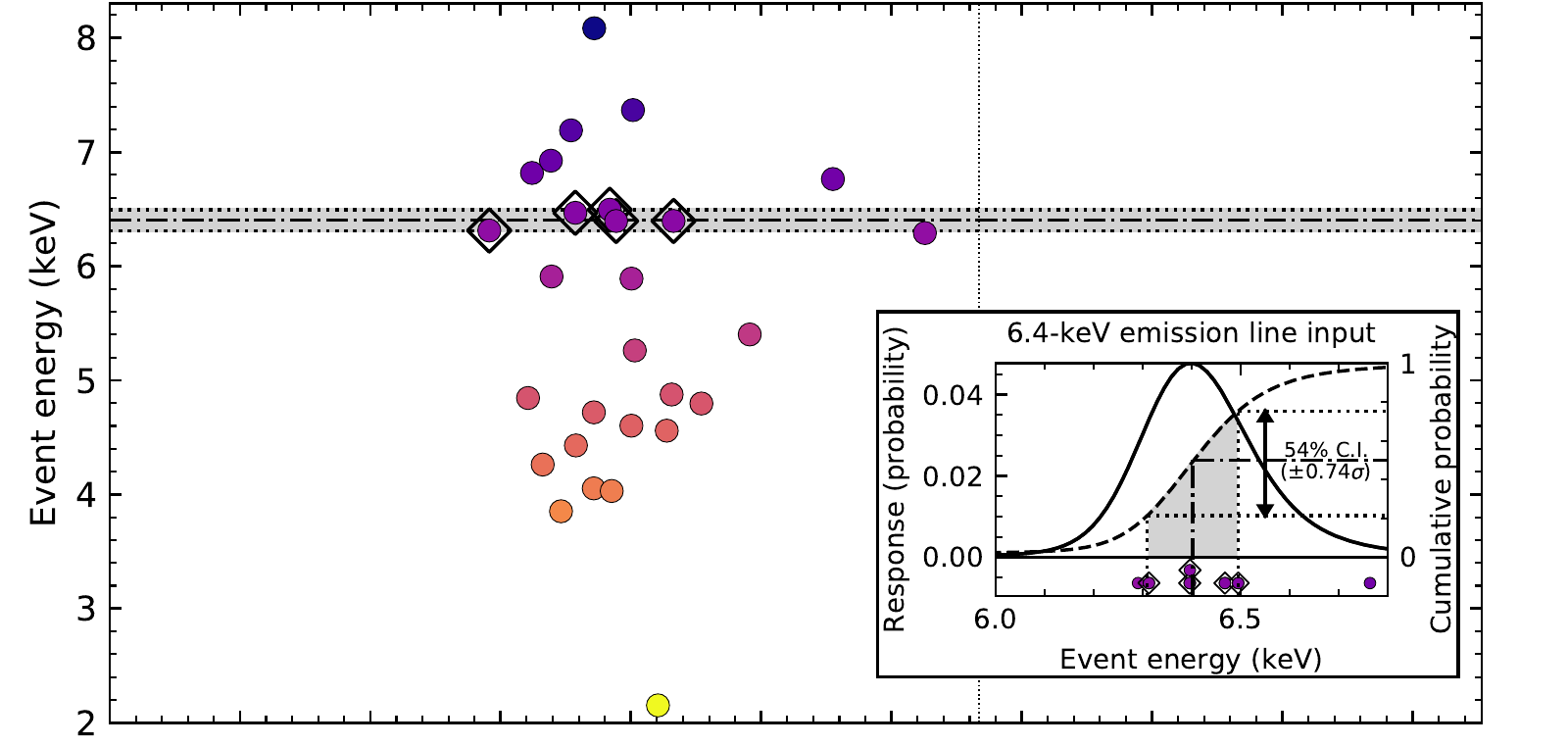}\\
                          &                                                                                     & \raisebox{35mm}{\bf c} &\includegraphics[width=80mm]{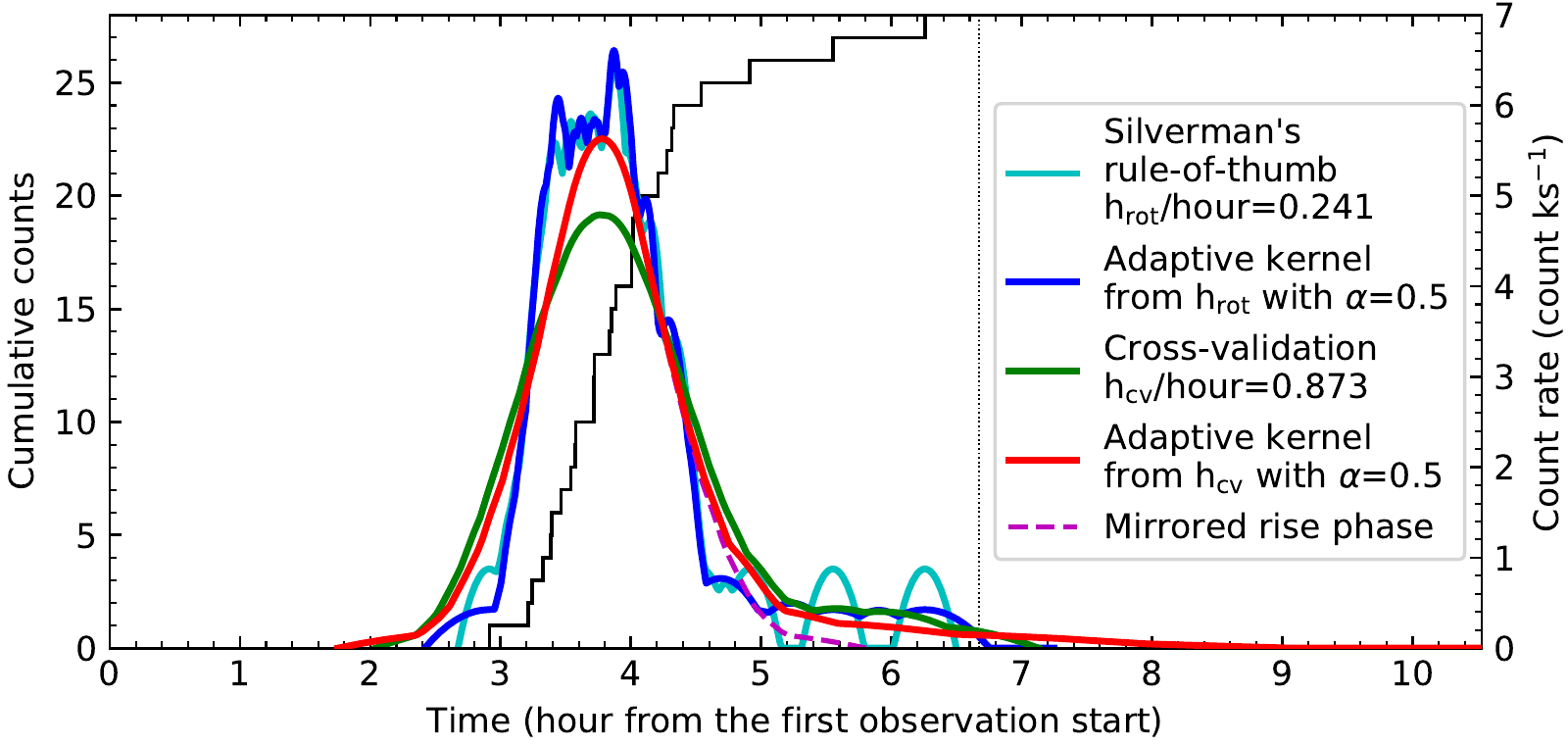}\\
\end{tabular}
\caption{X-ray flare of HOPS~383. 
\emph{Panel a}: the sky position of the X-ray events detected during the 
\emph{Chandra} observation on 2017 December 13. 
The cyan cross and circle are the X-ray position and positional 
error of HOPS~383, respectively. 
The dashed and solid contours encompass 90\% (position determination) 
and 96\% (spectrum extraction) of 1.49~keV
point source emission, respectively. 
The diamonds mark the events from the 6.4~keV emission line.
\emph{Panel b}: event energy versus time of arrival. 
The gray stripe is centered at 6.4~keV
and corresponds to a confidence interval of 54\% centered at this emission line 
observed with ACIS-I (inset).
\emph{Panel c}: cumulative counts and various smoothed light curves.
The dashed curve is the mirrored rise phase of the red curve.
\emph{Panels b and c}: the vertical dotted line 
at $\sim$6.7~h
marks an exposure gap of 3.1~s.
}
\label{fig:lc}
\end{figure*}

\section{X-ray results}

\subsection{X-ray flare of HOPS~383}

We detect a hard ($E>2$~keV) X-ray source, \object{CXOU~J053529.78-045950.4} 
(hereafter, CXOU~J053529),
that is spatially coincident with JVLA-NW (Fig.~\ref{fig:fov} 
and Appendix~\ref{appendix:cxo_astrometry}). 
CXOU~J053529 was not detected
in a \emph{Chandra} observation of the OMC-2 and 3 region
obtained in 2000 January at an off-axis 
angle of $6\farcm3$ \citep{tsujimoto02,getman17}.

This spatially unresolved X-ray source is only detected, with 28~counts in the 0.5--9.9~keV energy range, 
during the first \emph{Chandra} exposure that we obtained in 2017
(Fig.~\ref{fig:lc}a).
The first photon is detected $\sim$2.9~h 
after the start of the observation
and the X-ray burst
duration is $\sim$3.3~h (Fig.~\ref{fig:lc}b). 
We estimated the shape of the light curve 
by smoothing the event times of arrival
(Appendix~\ref{appendix:cxo_timing}).
The count rate peaks to $\sim$5.6~counts~ks$^{-1}$ 
in $\sim$0.9~h after the first photon detection, 
and it decays gradually in $\sim$2.5~h until the last photon detection
(red curve in Fig.~\ref{fig:lc}c).
This time variation, with rapid rise and slow decay, is similar to the typical
light curves of magnetic flares from young stellar objects \citep[YSOs;][]{getman08}.
X-rays may also be emitted by accretion shocks at the protostellar surface 
and/or shocks in outflows \citep{guedel09}, 
but the former produces plasmas that are not bright and hot
enough to emit the observed hard X-rays 
and the latter is constant on week timescales \citep{kastner05}, 
which is much longer than the hour variability that we observed.  
Therefore, this burst of X-rays indicates magnetic activity
from HOPS~383 that is associated with JVLA-NW.

\subsection{X-ray flare spectrum}

\begin{figure*}[!t]
\centering
\begin{tabular}{@{}l@{}l@{}l@{}l@{}}
\raisebox{72mm}{\bf a} & \includegraphics[width=75mm,]{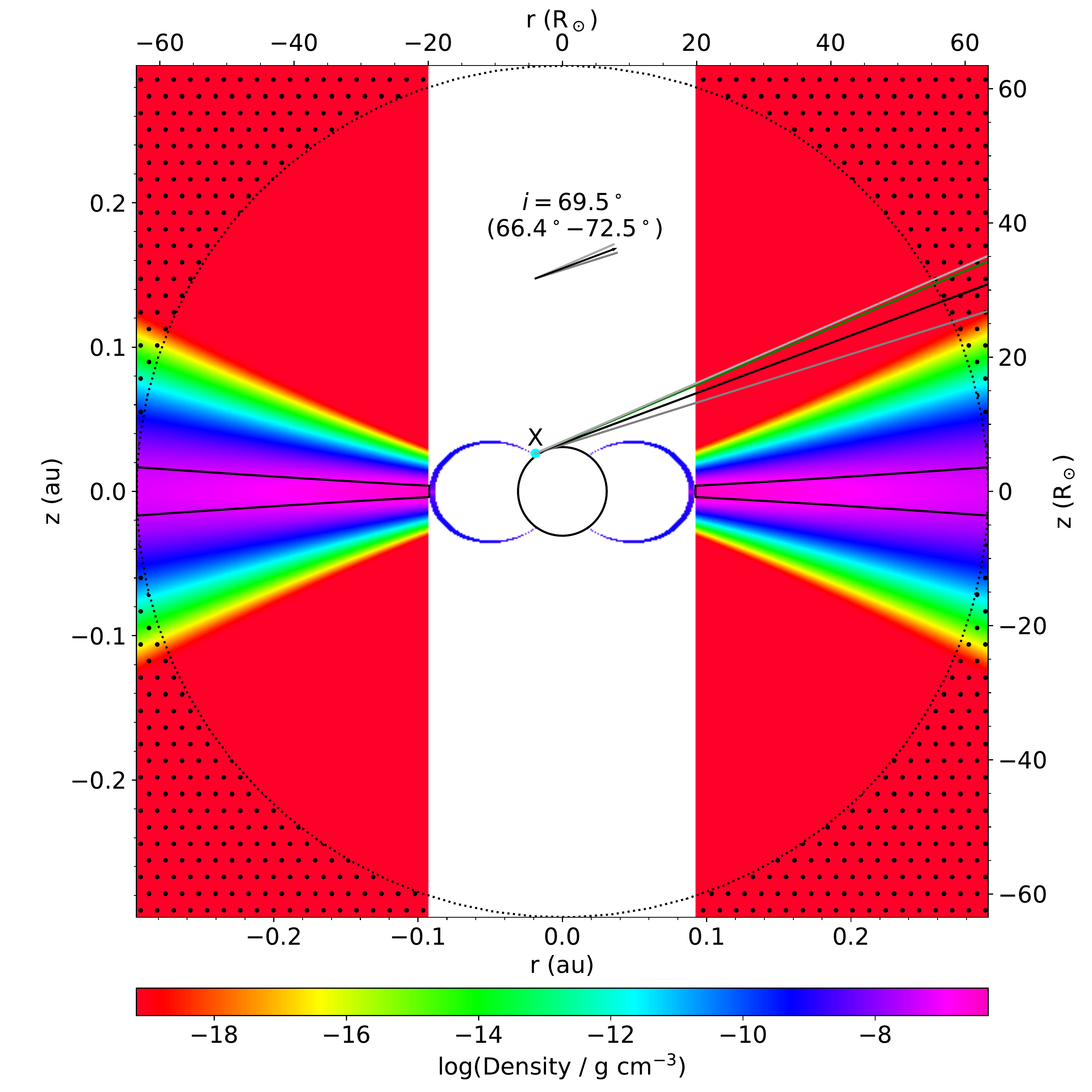} 
& \raisebox{72mm}{\bf b} & \includegraphics[width=75mm]{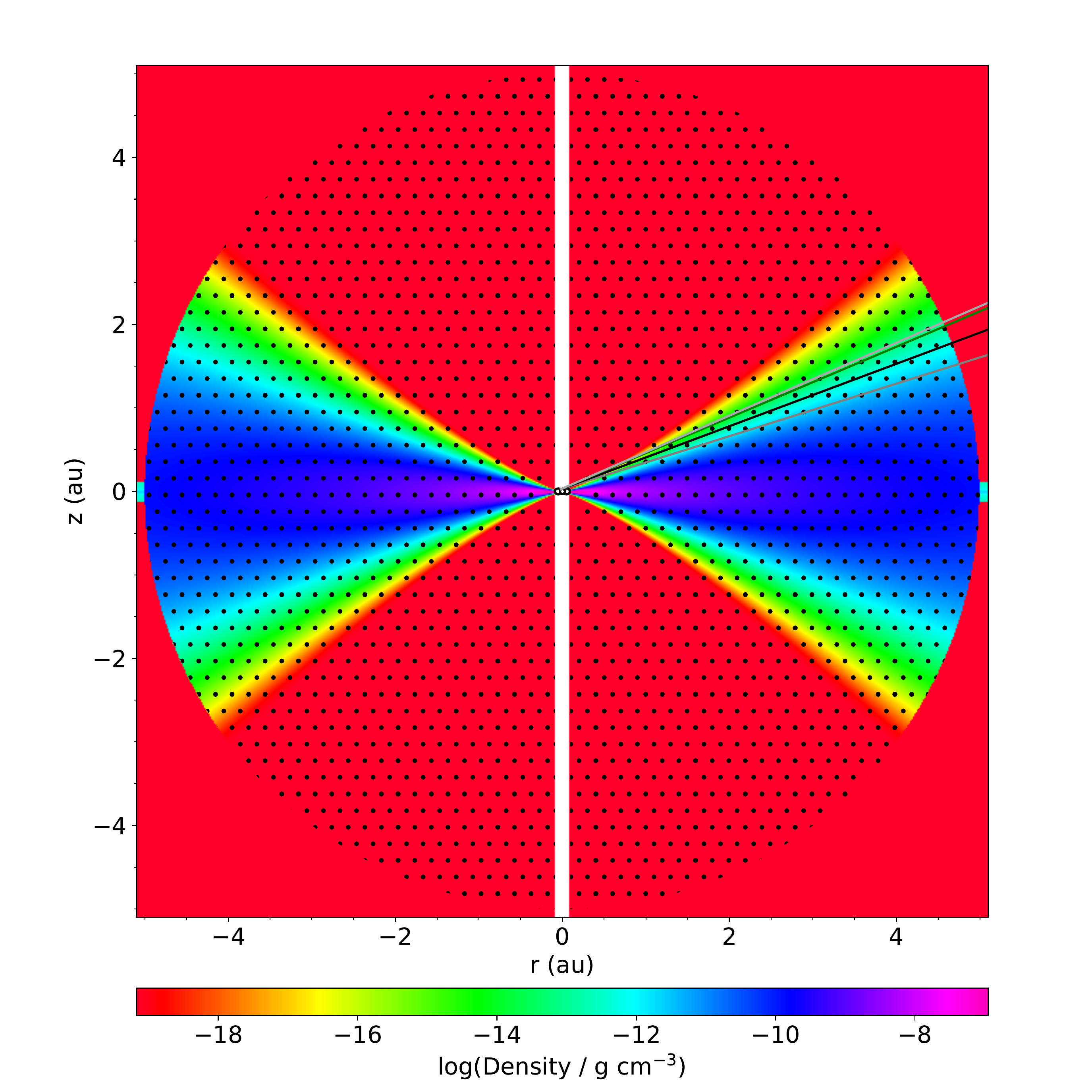} \\  
\raisebox{72mm}{\bf c} & \includegraphics[width=75mm]{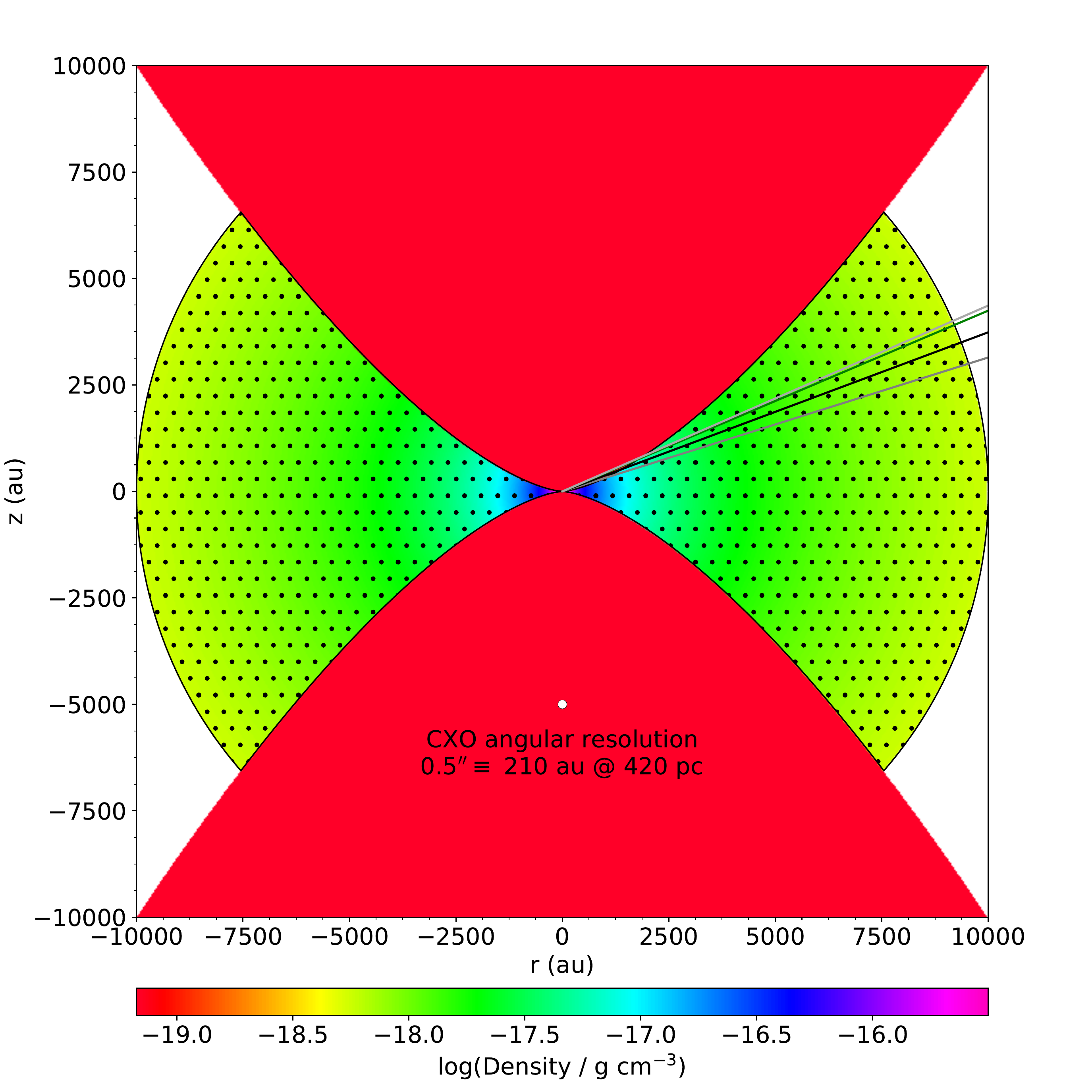}  
& \raisebox{72mm}{\bf d} & \raisebox{4mm}{\includegraphics[width=70mm]{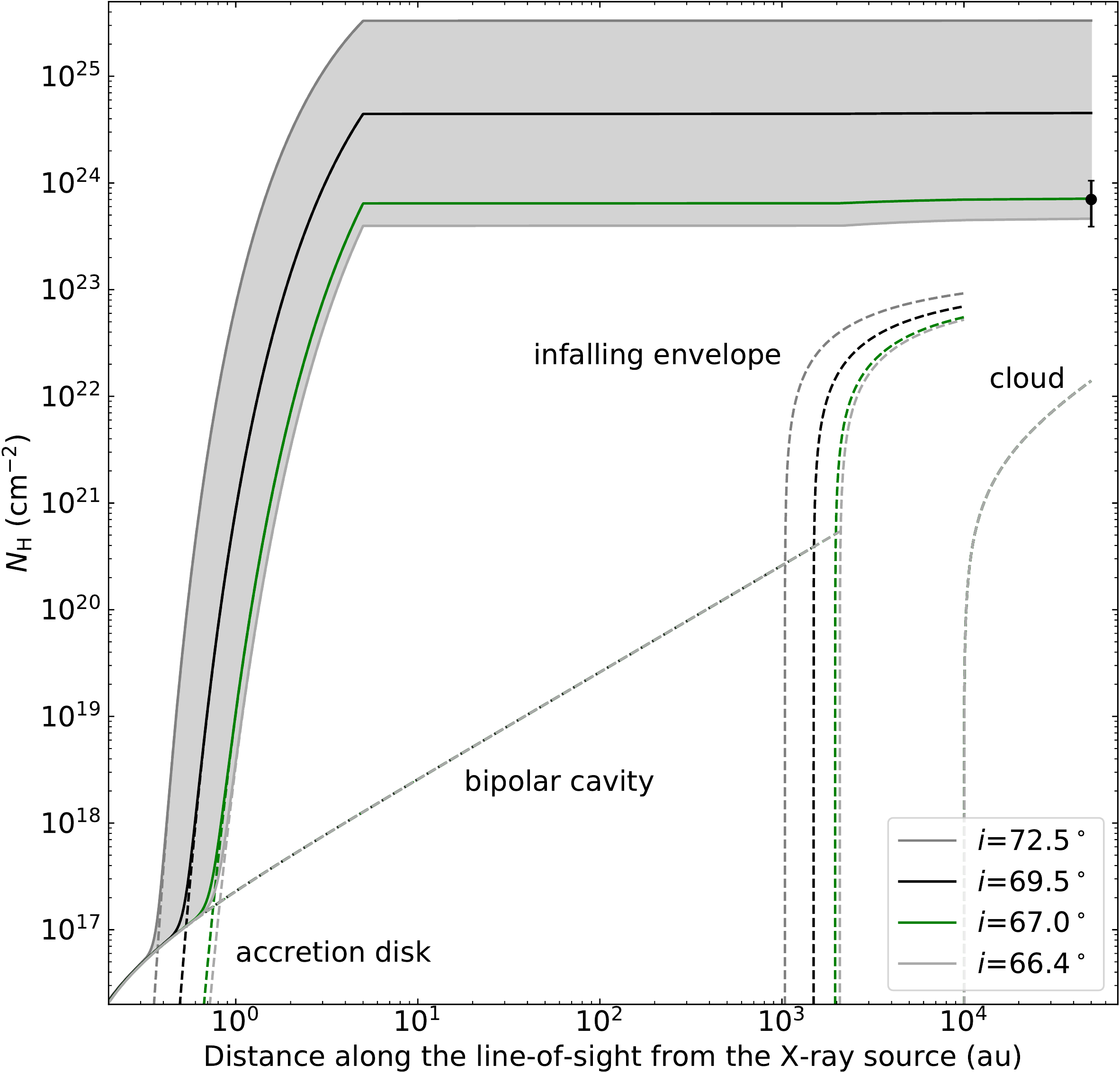}}
\end{tabular}
\vspace{-0.1cm}
\caption{Probing the gas and dust toward HOPS~383 with the X-ray spectrum. 
\emph{Panels a-c}: 2D density distribution 
from small to large spatial scales
and viewing angle
from the model of HOPS~383 \citep{furlan16}.
Dots indicate regions with dust.
\emph{Panel d}: cumulative hydrogen column density along the line-of-sight 
from the X-ray source.
The dashed lines show the contribution of the accretion disk, the bipolar cavity, 
the infalling envelope, and the cloud.
The data point with an error bar (90\% C.I.) is the hydrogen column density that we obtained 
from the X-ray spectrum modeling.
}
\label{fig:nh}
\end{figure*}

We modeled the X-ray spectrum with 
isothermal coronal plasma
emission
suffering from intervening absorption (Appendix~\ref{appendix:cxo_spectral}).
The best-fit model has an excess 
and a deficit
relative to the observed spectrum below 
and above 6.4~keV, respectively,
suggesting an additional component at 6.4~keV.
This deviation is likely associated with
the emission line arising from 
neutral or low-ionization iron.
Therefore, we 
included
in the model 
a Gaussian emission line centered at 6.4~keV.
We performed a Markov chain Monte Carlo analysis 
(MCMC)
to compute 
the probability density functions of the physical parameters,
from which we determined the median parameter
values and 90\% confidence intervals.
We find that CXOU~J053529 is 
highly embedded
with a hydrogen column density 
($N_\mathrm{H, X}$)
of $7.0^{+3.5}_{-3.1}\times10^{23}$~cm$^{-2}$.
This result for $N_\mathrm{H}$ is consistent with the large median energy 
of the 0.5--8 keV counts (MedE=5.402~keV)\footnote{ 
For epoch 2017, 
the degradation of the ACIS effective area at low energy due to the contamination 
of the optical blocking filter likely does not significantly modify 
the median energy of a source as hard as HOPS~383.}, 
given the $\log{N_\mathrm{H}}$ versus\ MedE relations found for YSOs in the Orion nebula cluster 
\citep[ONC; Fig.~8 of][]{feigelson05} and M17 \citep[Fig.~10a of][]{getman10}.

We detected the 6.4~keV emission line with a high probability of 98.9\%.
Our spectral analysis suggests that we have detected
five photons from the Fe~6.4~keV emission line;
these photons are likely the closest in energy to 6.4~keV.
Indeed, we observed five photons with energies
that are densely packed around the predicted 
emission feature at 6.4~keV
(diamonds in Fig.~\ref{fig:lc}b inset). 
Moreover, the spatial distribution of these Fe line photons is point-like 
(Fig.~\ref{fig:lc}a), suggesting 
all of the observed Fe fluorescence emission arises 
from within the \emph{Chandra} point spread function,
that is to say, a radius from HOPS~383 
that is lower than 105~au at the assumed distance 
to the protostar.
The iron line
equivalent width, $EW=1.1^{+2.1}_{-0.9}$~keV,
is large (even at the lower limit) relative to what was expected from the possible emission processes. 
The neutral iron can be ionized by photons 
or electrons that have energies larger than 
the Fe~K-shell ionization 
potential of 7.112~keV. 
The contributions of these two ionization channels to the observed X-ray 
line in accreting young stars 
and Class~I protostars \citep{guedel09,hamaguchi05,czesla10,hamaguchi10,pillitteri19}
are not firmly settled,  
but photoionization is less energetically challenging \citep{czesla10}. 
The plasma temperature is not well constrained  
mainly due to the low count rate, but a hot plasma temperature is favored
($kT=4.1^{+13.8}_{-2.8}$~keV),
which is consistent with a magnetic flare and 
the photoionization of iron.

\subsection{X-ray flare energetics}

The mean value of the absorption-corrected luminosity during
the flare time interval is
$1.7^{+5.5}_{-1.1}\times10^{31}$~erg~s$^{-1}$
in the 2--8~keV energy band. 
Multiplying this value by the flare peak to flare mean count rate ratio (2.39) 
provides the flare peak value of $\sim$$4.2^{+13.0}_{-2.6}\times10^{31}$~erg~s$^{-1}$.
For comparison purposes, the maximum peak value of the 84 X-ray flares with typical 
shapes observed from YSOs in the ONC
\citep{getman08} is $8.3\times10^{32}$~erg~s$^{-1}$ in the 0.5--8~keV energy band; 
only $\sim$34\% of these flares have peak values that are higher than the HOPS~383 one.
During our observations, the non-detection of HOPS~383 outside the flare time interval implies that 
its quiescent absorption-corrected luminosity is lower than 
$2.0\times10^{30}$~erg~s$^{-1}$ 
in the 2--8~keV energy band (Appendix~\ref{appendix:quiescent}).
Therefore, the flare that we detected peaked at least $21^{+65}_{-13}$
times above the quiescent level.
The lower limit on
the total energy released in X-rays during this flare, as 
computed from the mean luminosity times the flare duration, 
was $2.0^{+6.6}_{-1.3}\times10^{35}$~ergs.

\section{Discussion}

Starting from a model of HOPS~383 consisting of
an infalling envelope with bipolar cavities carved 
by outflows and a small (5~au radius) accretion disk 
(Fig.~\ref{fig:nh} a--c),
constrained by the post-outburst spectral energy distribution (SED)
using 
a
Monte Carlo radiative transfer 
code \citep{furlan16}, we computed the predicted $N_\mathrm{H}$ toward the central 
protostar (Appendix~\ref{appendix:model}).
Due to the high inclination angle in this model, 
the line-of-sight intercepts the upper layers of 
the accretion disk, which produce the bulk of $N_\mathrm{H}$;
moreover, this value is extremely sensitive to 
the viewing angle (Fig.~\ref{fig:nh}d).
The observed $N_\mathrm{H, X}$ favors a slightly lower inclination.
However, the available grid of 
protostellar models \citep{furlan16} does
not include a lower-mass accretion disk
or a model with no accretion disk at all, which may reproduce both the SED and 
$N_\mathrm{H, X}$.
These model trials illustrate the importance of X-ray 
constraints on $N_\mathrm{H}$ to
the overall consistency of protostar modeling.

In the optically thin case at 6.4~keV, that is to say, as long as the H
column density by a cold absorber with a solar elemental abundance 
($N_\mathrm{H}^{\prime}$) is 
lower than $\sim$$10^{24}$~cm$^{-2}$,
the iron line
$EW$ produced by photoionization is 
$EW = 2.5\,(\Omega/4\pi)(N_\mathrm{H}^{\prime} / 10^{22}$~cm$^{-2})$~eV, 
where $\Omega$ is the subtended angle of the irradiated material 
from the X-ray irradiating source \citep{tsujimoto05}.
Since $\Omega/4\pi < 1$, the large observed value of $EW$ in HOPS~383 
rules out the optically thin condition.
In the optically thick case of the photoionization 
of photospheric iron by an X-ray flare,
the iron line
$EW$ that is computed by Monte-Carlo radiative transfer simulations 
is limited to $\sim$130~eV for solar iron abundance 
and can only be increased by 
factors $<2$ by disk flaring \citep{drake08}.
However, large iron line
$EW$ can be obtained in both optical depth 
conditions when the photoionizing X-ray source
is partly eclipsed \citep{drake08}.
In the model, the X-ray source is 
located on the obscured region of the star
and
near the base of the accretion funnels (Fig.~\ref{fig:nh}a).
This source irradiates the accretion funnels, 
the upper layers of the accretion disk (Fig.~\ref{fig:nh}b),
and the circumstellar envelope (Fig.~\ref{fig:nh}c).

\section{Conclusions}

Our detection of an X-ray flare from HOPS~383 provides
conclusive evidence that 
strong magnetic activity is present 
at this
bona fide Class~0 protostar.
The resulting
X-ray irradiation contributes to the ionization 
of the base of the outflow \citep{shang04},
and the magnetic reconnection that triggers 
powerful flares, 
similar to
the one we detected,
most likely drives energetic particle ejections \citep{feigelson02}.
Such protostellar cosmic rays have been proposed
to collide with refractory dust grains located 
at the inner edge of the accretion disk, inducing spallation reactions 
that could yield short-lived radionuclei, such as are observed
in the refractory inclusions of chondritic meteorites \citep{gounelle13,sossi17}.
Mass ejections have similarly
been invoked in the 
case of the bona~fide Class~0 protostar
\object{OMC-2~FIR~4} to explain the production 
of free electrons via collisions in the envelope, 
thereby enhancing the abundances of molecular ions \citep{ceccarelli14}.
The Class~0 stage ($\sim$$10\,000$~yr) 
has an $\sim$10 times shorter 
duration than the Class~I stage;
however, during the earliest stage, half of the envelope mass is accreted, 
building the central star and the accretion disk. Therefore, the 
determination of the internal irradiation level 
in Class~0 protostars is paramount
for the understanding of protostellar chemistry.

Longer observations are required to determine the 
flaring activity level and to collect more photons from such 
deeply-embedded, nascent stars.
Athena X-IFU \citep{barret18}, which is scheduled to
be launched at the beginning of the 2030s,
will improve the X-ray throughput and spectral resolution at 6.4~keV,
enabling a potential leap in our knowledge of the 
onset of protostellar magnetic activity.


\begin{acknowledgements}
We thank the anonymous referee for the careful reading of the manuscript.
We are grateful to Joshua Wing (\emph{Chandra} X-ray Center; CXC) 
and Steve Heathcote (Director of the Cerro Tololo Inter-American Observatory) 
for the scheduling of the simultaneous \emph{Chandra} and SOAR observations.
K.H., D.P., and N.G.\ 
are thankful to Jonathan (Jay) Elias (SOAR Director) and 
Patricio Ugarte (Observer Support) 
for technical support during 
the Spartan observation. 
N.G.\ thanks Patrick~Broos and Leisa~Townsley for 
Acis~Extract support.
This work was supported by ``the Programme National de Physique Stellaire'' (PNPS) 
of CNRS/INSU co-funded by CEA and CNES.
Support for this work was provided by the National Aeronautics and Space 
Administration (NASA) through \emph{Chandra} Award Numbers 
GO7-18012A and B
issued by the 
CXC, which is operated by the Smithsonian Astrophysical Observatory for and 
on behalf of the NASA under contract 
NAS8-03060.
The scientific results reported in this article are based on observations made by the 
\emph{Chandra} X-ray Observatory (ObsIDs 18927, 20882 and 20883).
This research has made use of software provided by the 
CXC in the application packages CIAO.
SOAR is a joint project of the Minist\'erio da Ci\^encia, Tecnologia, e Inova\c{c}\~ao
(MCTI) da Rep\'ublica Federativa do Brasil, the U.S. National Optical Astronomy
Observatory (NOAO), the University of North Carolina at Chapel Hill (UNC),
and Michigan State University (MSU).
This publication makes use of data products from 
WISE,
which is a joint project of the University of California (Los Angeles) 
and the Jet Propulsion Laboratory/California Institute of Technology (JPL/Caltech), 
and NEOWISE
which is a project of the JPL/Caltech. 
WISE and NEOWISE are funded by the NASA.
This work has made use of data from the European Space Agency (ESA) 
mission \emph{Gaia}
(\href{https://www.cosmos.esa.int/web/gaia}{https://www.cosmos.esa.int/web/gaia}, 
processed by the \emph{Gaia} Data Processing and Analysis Consortium 
(DPAC, \href{https://www.cosmos.esa.int/web/gaia/dpac/consortium}{https://www.cosmos.esa.int/web/gaia/dpac/consortium}). 
Funding for the DPAC has been provided by national institutions, 
in particular the institutions
participating in the \emph{Gaia} Multilateral Agreement.
This research has made use of Python and astropy, corner, matplotlib, numpy, pyregion, scipy and statsmodels modules.
\end{acknowledgements}


\phantomsection
\addcontentsline{toc}{chapter}{References}



\phantomsection
\addcontentsline{toc}{chapter}{Appendices}
\begin{appendix}


\section{Protostellar evolutionary stage}
\label{appendix:stage}

\begin{figure}[!ht]
\centering
\includegraphics[width=90mm]{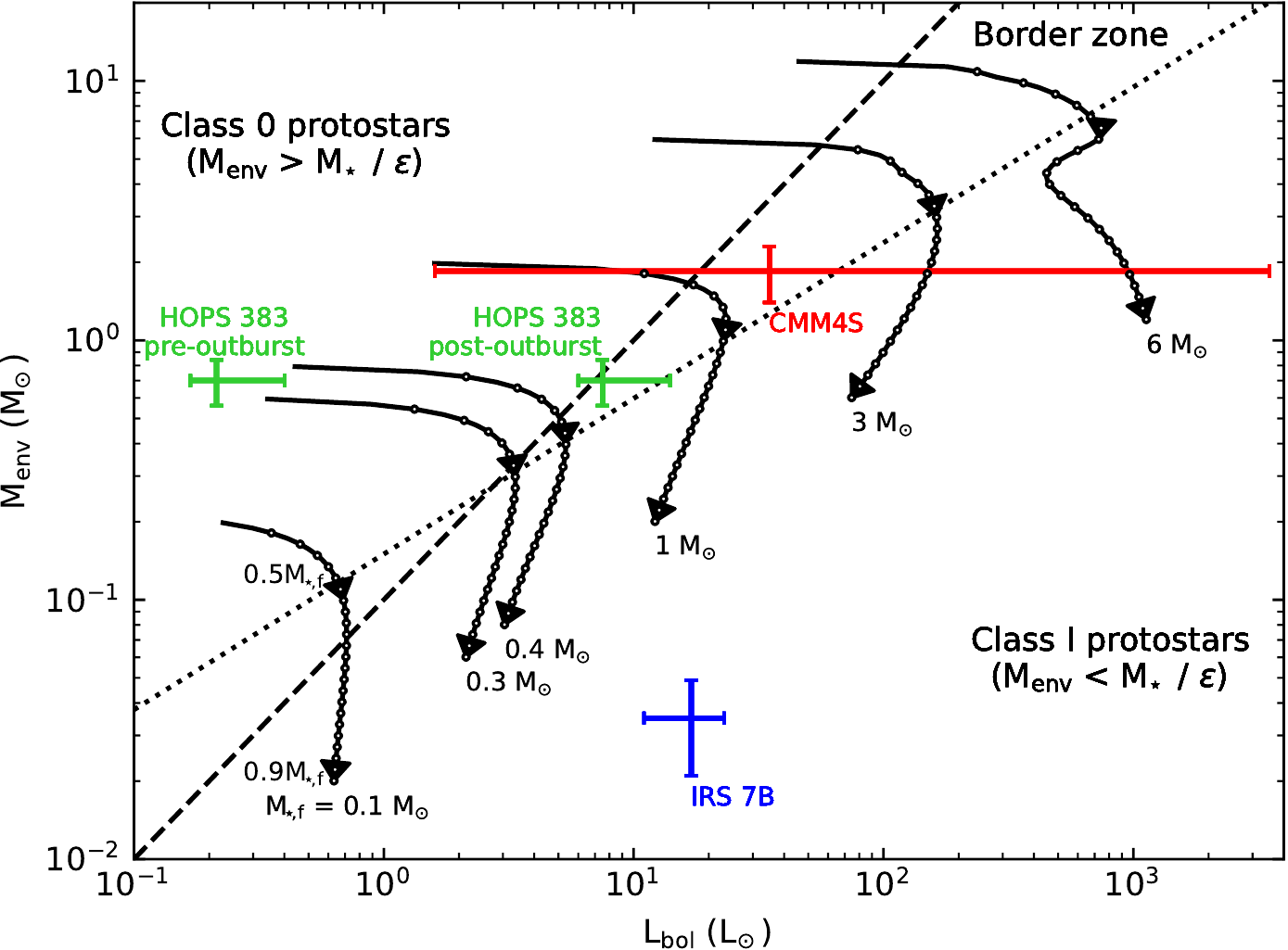}
\caption{Envelope mass versus bolometric-luminosity diagram comparing 
the pre- and post-outburst positions of HOPS~383 
(green data)
with previously reported 
Class~0/I 
low-mass
protostars 
detected in X-rays
(blue and red data).
The border zone between Class~0 ($M_\mathrm{env} > M_\star/\epsilon$,
where $\epsilon=0.5$ is the local star-formation efficiency) 
and Class~I ($M_\mathrm{env} < M_\star/\epsilon$) protostars is defined 
by $M_\mathrm{env} = 0.1$~$L_\mathrm{bol}$ (constant accretion; dotted line) 
and $M_\mathrm{env} = 0.15~{L_\mathrm{bol}}^{0.6}$ (exponentially decaying accretion; dashed line).
The protostellar evolutionary tracks are shown from 0.01$M_\mathrm{\star,f}$ to 0.9$M_\mathrm{\star,f}$, 
where $M_\mathrm{\star,f}$ is the final stellar mass.
The rising time along the tracks is indicated by white dots every 
$10\,000$~yr. 
}
\label{appendix_fig:diagram}
\end{figure}

Figure~\ref{appendix_fig:diagram} shows
the envelope mass ($M_\mathrm{env}$) versus bolometric-luminosity 
($L_\mathrm{bol}$) diagram, 
which is used
as an evolutionary diagnostic
\citep{saraceno96,andre00,andre08}. 
In this scenario, the Class~0 and I border zone is defined 
by $M_\mathrm{env} = 0.1$~$L_\mathrm{bol}$ \citep[constant accretion, see][]{andre94} 
and $M_\mathrm{env} = 0.15~{L_\mathrm{bol}}^{0.6}$ 
\citep[exponentially decaying accretion, see][]{bontemps96}.
Propagating the model and flux uncertainties of HOPS~383 \citep{safron15}, 
we 
gather that:
$M_\mathrm{env}=0.70\pm0.14$~$M_\odot$, 
$L_\mathrm{bol}^\mathrm{post{\mhyphen}outburst}=7.5^{+6.5}_{-1.5}$~$L_\odot$,
and $L_\mathrm{bol}^\mathrm{pre{\mhyphen}outburst}=0.21^{+0.19}_{-0.05}$~$L_\odot$ 
(obtained from the preoutburst-to-outburst flux ratio at 24~microns of $35\pm3$).
The post-outburst position of HOPS~383 is within this Class~0 and I border zone due to 
its high accretion luminosity, but the pre-outburst position of HOPS~383 is well above it, 
where $M_\mathrm{env}$ is larger than $M_\star$, confirming 
its
bona fide Class~0 stage.
For comparison purposes, 
the infrared source
\object{IRS~7B} 
in the R~Coronae
Australis star-forming core, which has a deeply embedded
($N_\mathrm{H} \sim 3\times10^{23}$~cm$^{-2}$) 
 X-ray counterpart 
that was proposed as a Class~0 protostar candidate 
\citep{hamaguchi05}, is
on the side of the Class~I stage in this plot,
based on the bolometric luminosity 
reported in \cite{groppi07}. 
The bolometric luminosity of the Class~0 candidate \object{CMMS4} \citep{kamezaki14} 
is not accurate enough for a robust conclusion.

We computed the protostellar evolutionary tracks assuming the following
\citep{bontemps96,andre00,andre08}: 
an initial envelope mass, $M_\mathrm{env}(0)$,
decays exponentially with a characteristic timescale, $\tau = 10^5$~yr; 
a mass accretion rate, $\dot{M}_\mathrm{acc}(t)=\epsilon M_\mathrm{env}(t) / \tau$,
where the local star-formation efficiency is $\epsilon=0.5$ 
\citep{andre08}; and 
$L_\mathrm{bol}(t)=L_\mathrm{acc}(t)+L_\star(t)$, 
where $L_\mathrm{acc}(t)=G \dot{M}_\mathrm{acc}(t)M_\star(t)/R_\star(t)$ 
is the accretion luminosity, 
where $G$ is the gravitational constant, and $M_\star$(t), $R_\star(t)$, and $L_\star(t)$ 
are the protostellar mass, radius, 
and interior stellar luminosity, respectively, on the birthline of Fig.~1 
in \cite{palla99} 
where a constant accretion rate of $10^{-5}$~$M_\odot$~yr$^{-1}$ is assumed 
to compute the radius-mass relation \citep{stahler88,palla92}.
We estimate with this toy model that the final stellar mass of HOPS~383 will be 
$\sim$$0.4M_\odot$ and that less than $\sim$10\% of it has already been accreted onto 
the central protostar.


\section{WISE and NEOWISE-R photometry}
\label{appendix:wise}

\begin{figure}[!h]
\centering
\includegraphics[width=90mm]{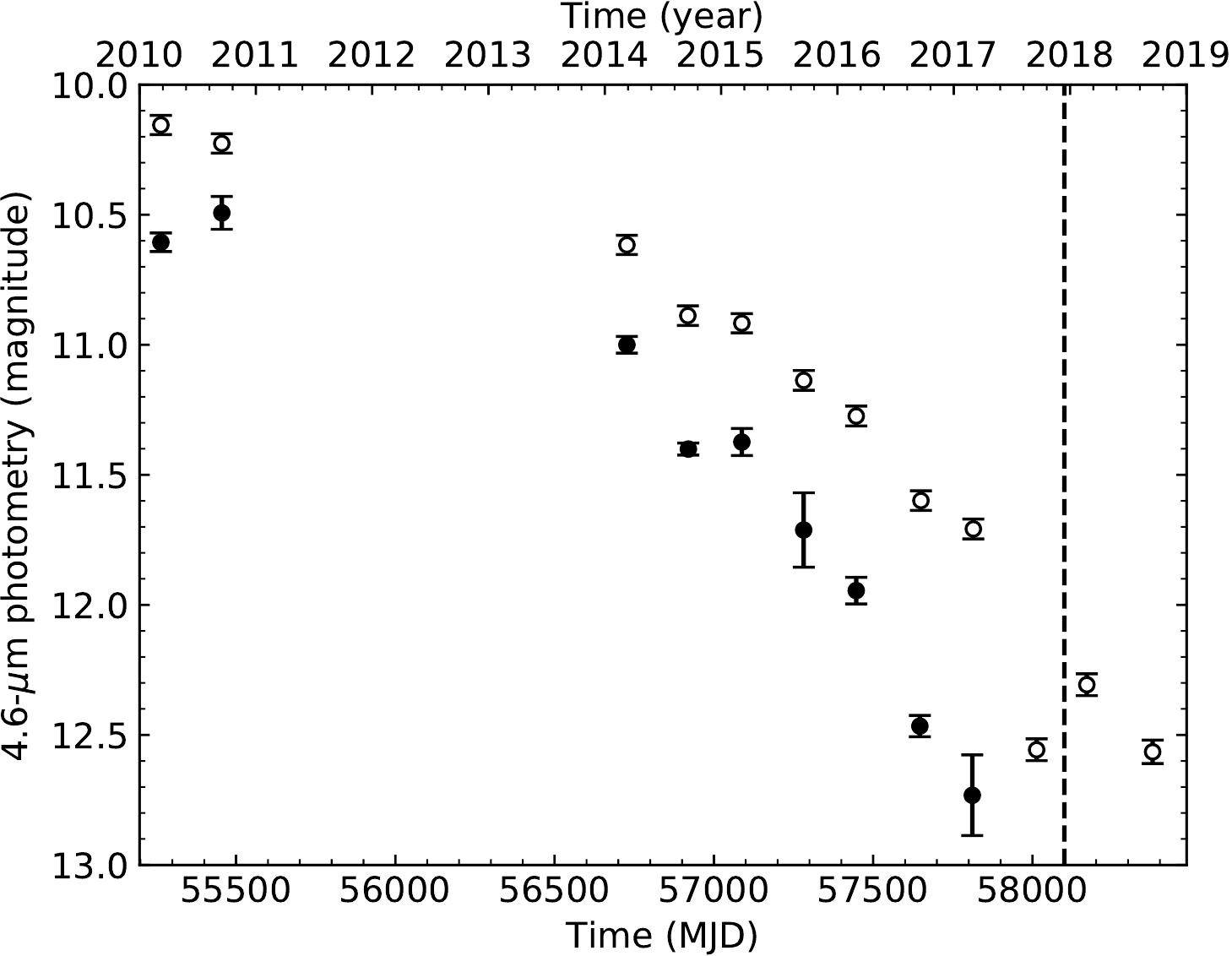}
\caption{Decay of the episodic accretion of HOPS~383 in the mid-infrared.
The 2010 and 2014--2018 data points are from WISE and NEOWISE observations
with the W2 filter at 4.6~$\mu$m, respectively. 
The black and white dots are the profile-fit 
(Table~\ref{appendix_tab:wise})
and aperture 
(Table~\ref{appendix_tab:wise_aper})
photometry 
for each observing epoch, respectively. 
The dashed vertical line is the arrival time of the first 
detected
photon of the X-ray flare.
}
\label{appendix_fig:midir_light_curve}
\end{figure}

\begin{table}[!h]
\caption{WISE and NEOWISE-R 4.6-$\mu$m 
profile-fit
photometry of HOPS~383.}
\label{appendix_tab:wise}
\centering
\smallskip
\begin{tabular}{@{}lccccc@{}}
\hline\hline\noalign{\vskip 1.2mm}      
Instrument &  $\langle$MJD$\rangle$ & $\Delta$MJD & N & $\langle$w2mpro$\rangle$ & $\sigma$\\
           &                        & (day)          &                     & (mag)                      & (mag)\\
\noalign{\vskip 1.2mm}\hline\noalign{\vskip 1.2mm}
     WISE  & 55\,263.89 & 0.99 &  6 & 10.606 & 0.036 \\
     WISE  & 55\,455.31 & 0.73 &  6 & 10.493 & 0.063 \\
NEOWISE-R  & 56\,726.88 & 4.54 &  8 & 11.000 & 0.032 \\
NEOWISE-R  & 56\,919.85 & 0.72 &  7 & 11.401 & 0.023 \\
NEOWISE-R  & 57\,087.73 & 0.99 & 10 & 11.374 & 0.052 \\
NEOWISE-R  & 57\,281.88 & 1.12 &  8 & 11.712 & 0.143 \\
NEOWISE-R  & 57\,446.82 & 0.98 &  7 & 11.945 & 0.051 \\
NEOWISE-R  & 57\,646.54 & 0.13 &  2 & 12.466 & 0.041 \\
NEOWISE-R  & 57\,811.16 & 0.26 &  2 & 12.732 & 0.155 \\
\hline
\end{tabular}
\tablefoot{The profile-fit photometry was computed from the single-exposure (L1b) source tables
using the mean and standard deviation of the N profile-fit fluxes 
in the W2 band (w2mpro) obtained during each observing epoch. We note that
$\Delta$MJD is the time interval between the first and the last frame 
for each epoch.}
\end{table}

\begin{table*}[!t]
\caption{WISE and NEOWISE-R 4.6-$\mu$m 
aperture 
photometry of HOPS~383.}
\label{appendix_tab:wise_aper}
\centering
\begin{tabular}{@{}lccccccc@{}}
\hline\hline\noalign{\vskip 1.2mm} 
Instrument &  $\langle$MJD$\rangle$ & $\Delta$MJD & N & MAGZP & w2mcor & $\langle$w2mag$\rangle$ & $\sigma$\\
           &                        & (day)          &                     & (mag)  & (mag)  & (mag)                      & (mag)\\
\noalign{\vskip 1.2mm}\hline\noalign{\vskip 1.2mm} 
     WISE  & 55\,263.89 & 0.99 & 13 & 19.596 & 0.331 & 10.155 & 0.037 \\
     WISE  & 55\,455.11 & 1.12 & 15 & 19.692 & 0.331 & 10.226 & 0.037 \\
NEOWISE-R  & 56\,726.88 & 4.54 & 10 & 19.669 & 0.322 & 10.616 & 0.037 \\
NEOWISE-R  & 56\,918.17 & 4.87 & 16 & 19.642 & 0.322 & 10.888 & 0.038 \\
NEOWISE-R  & 57\,087.66 & 1.12 & 14 & 19.645 & 0.326 & 10.917 & 0.037 \\
NEOWISE-R  & 57\,281.94 & 1.25 & 16 & 19.641 & 0.326 & 11.137 & 0.038 \\
NEOWISE-R  & 57\,446.82 & 0.98 &  7 & 19.641 & 0.326 & 11.274 & 0.038 \\
NEOWISE-R  & 57\,649.88 & 8.78 & 17 & 19.644 & 0.326 & 11.599 & 0.038 \\
NEOWISE-R  & 57\,814.43 & 7.99 & 15 & 19.644 & 0.326 & 11.708 & 0.038 \\
NEOWISE-R  & 58\,013.16 & 1.11 & 16 & 19.643 & 0.326 & 12.557 & 0.042 \\
NEOWISE-R  & 58\,170.74 & 6.48 & 12 & 19.643 & 0.326 & 12.307 & 0.042 \\
NEOWISE-R  & 58\,377.31 & 1.11 & 13 & 19.643 & 0.326 & 12.565 & 0.045 \\
\hline
\end{tabular}
\tablefoot{The aperture photometry in each observing epoch was computed from coadded 
single exposures of a good quality. The aperture correction (w2mcor) 
is taken from the single exposure (L1b) source tables.
The uncertainty of the magnitude zero-point (MAGZP) is fixed to 0.037~mag.}
\end{table*}

The mid-infrared flux of HOPS~383 started
to rise between 2004 and 2006, 
peaked by 2008 with a 24~$\mu$m flux that was
35 times brighter than the pre-outburst flux,
and showed no large decay between 2009 and 2012 \citep{safron15}.
The sky position of HOPS~383 was scanned in mid-infrared twice per year 
by the Wide-field Infrared Survey Explorer (WISE) mission \citep{wright10b} in 2010 
and by the Near-Earth Object Wide-field Infrared Survey Explorer (NEOWISE) 
reactivation mission \citep{mainzer14} 
in 2014--2019 for which several single exposures were obtained at each observing 
epoch\footnote{\href{https://irsa.ipac.caltech.edu/Missions/wise.html}{https://irsa.ipac.caltech.edu/Missions/wise.html}\label{note1}}. 

The All-Sky, WISE 3-band Cryo, and NEOWISE Reactivation 2015--2019 data releases 
provide single exposure (L1b) source tables for 2010 and 2014--2018.
The profile-fit photometry of HOPS~383 in the W2 band at 4.6~$\mu$m is available 
from 2010~March to 2017~February. 
As the full-width half-maximum (FWHM) in the W2 band is limited to $6\arcsec$, 
HOPS~383 lies on the wing of the point-spread-function of a brighter source 
(2MASS~J05353060-0459360), which is located at $15\farcs3$ northeast.
This profile fit computed on 1.5$\times$FWHM in radius mitigates this contamination.
For each observing epoch, 
we selected the best profile-fit fluxes ($w2rchi2 \le 2$) from the best single-exposure frames 
($qual\_frame > 0$, $qi\_fact > 0$, $saa\_sep > 0$, and $moon\_masked = 0$)
and we computed the mean and standard deviation of these fluxes.
The mean profile-fit flux decays by
2.1$\pm$0.2~mag in 7~years 
(black dots in Fig.~\ref{appendix_fig:midir_light_curve}
and Table~\ref{appendix_tab:wise}).

For each observing epoch, we selected the W2-band frames
of a good quality ($qual\_frame >0$) and 
stacked them with the WISE/NEOWISE 
coadder$^{\ref{note1}}$
using simple area weighting and 
a pixel scale of $1\farcs375$
for the final image. 
HOPS~383 is still visible after 2017~February in these stacked images as a faint source.
To obtain aperture photometry, we used a standard $8\farcs25$ radius aperture centered 
on the Spitzer position of HOPS~383 \citep{megeath12} 
and a custom $11\arcsec$ radius background-aperture located $22\arcsec$ west of it,
on a region free of point sources with lower extended emission 
compared to the $50\arcsec-70\arcsec$ radius annulus used in the single exposure source tables.
The aperture correction (w2mcor) is taken from the single exposure (L1b) source table.
The uncertainty of the magnitude zero-point (MAGZP) is fixed 
to the value of the WISE all-sky single exposures ($\sigma_\mathrm{MAGZP}=0.037$~mag).
The aperture flux decays by
2.40$\pm$0.06~mag in 7.5~years 
(white dots in Fig.~\ref{appendix_fig:midir_light_curve}
and Table~\ref{appendix_tab:wise_aper}), 
which is consistent with the behavior of the profile-fit flux, 
despite 
contaminating
sources in the source aperture.
The aperture flux increases by
0.25$\pm$0.06~mag between 2017~September and 2018~February, 
before coming back in 2018~September to the lower level that was observed previously.
We conclude that the decay of the 2008
accretion outburst ended by 2017~September, 
which was three months before our \emph{Chandra} observations (Table~\ref{appendix_table:cxo_log}).


\section{\emph{Chandra} ACIS-I data} 

\subsection{Observations}
\label{appendix:cxo_log}

The \emph{Chandra} observations are listed in Table~\ref{appendix_table:cxo_log}. 


\begin{table}[!h]
\caption{Log of \emph{Chandra} observations with ACIS-I 
(PI: N.G.).}
\label{appendix_table:cxo_log}
\centering
\begin{tabular}{cccc}
\hline\hline\noalign{\vskip 1.2mm}
ObsID & Start Time &  Exposure \\
           &        (UT)     &    (s)         \\
\noalign{\vskip 1.2mm}\hline\noalign{\vskip 1.2mm}
18927 & 2017-12-13T04:00:48 & 37\,411\\ 
20882 & 2017-12-14T01:24:04 & 32\,652\\ 
20883 & 2017-12-14T17:09:31 & 13\,814\\ 
\hline
\end{tabular}
\end{table}

\subsection{Data reduction}
\label{appendix:cxo_reduction}

We used the CIAO package\footnote{\href{https://cxc.cfa.harvard.edu/ciao/}{https://cxc.cfa.harvard.edu/ciao/}} 
\citep[version 4.9 with the calibration database version 
4.8.1;][]{fruscione06}
to reprocess the datasets 
(Table~\ref{appendix_table:cxo_log})
with the tool {\tt chandra\_repro}.
The source detection was performed in each observation with the CIAO tool {\tt wavdetect} 
with a detection threshold $1\times10^{-5}$   
and including information from the 
point spread function (PSF).
Twelve images were used per ObsID corresponding to three energy bands (0.5--2, 2--7, and 0.5--7~keV) 
and four resolution images (0\farcs25, 0\farcs5, 1\arcsec, and 2\arcsec)
that were searched for with {\tt wavdetect} on various pixel scales 
(1--2, 1--4, 1--8, and 1--16 pixel scales, 
corresponding to 0\farcs5--1\farcs5, 1\arcsec--8\arcsec, and 2\arcsec--32\arcsec, respectively)
to mitigate the spatial variation of the PSF on the detector \citep{broos10}.
The obtained {\tt wavdetect} source lists were then merged by pairs using 
the source matching  
IDL program {\tt match\_xy} 
in the Tools for ACIS Review \& Analysis (TARA) package\footnote{\href{http://personal.psu.edu/psb6/TARA/}{http://personal.psu.edu/psb6/TARA/}} 
\citep{broos10}.

The events were 
extracted using the ACIS Extract (AE) software 
package\footnote{\href{http://personal.psu.edu/psb6/TARA/ae_users_guide.html}{http://personal.psu.edu/psb6/TARA/ae\_users\_guide.html}} 
\citep[version ae2018june14;][]{broos10} 
 following the multi-pass validation procedure 
(revision 1099 of 2018-06-11)\footnote{\href{http://personal.psu.edu/psb6/TARA/procedures/}{http://personal.psu.edu/psb6/TARA/procedures/}}.
The AE source apertures were centered on source positions 
with a default PSF fraction of 90\% at 1.49~keV,
which was determined from point source simulations
computed by raytracing with MARX \citep{davis12}. 
The AE background regions were built to get enough background
counts in order to have the source photometry errors within 3\% of the values that would be obtained 
with no uncertainty in the background. 
New positions were determined for isolated sources with an off-axis position lower than 5$^\prime$
using the mean position of the extracted events. 
These positions were used to determine the astrometric position offsets between the ObsIDs 
and a reference catalog. We used the \emph{Gaia} data release~2 
(DR2)\footnote{\href{http://gea.esac.esa.int/archive/}{http://gea.esac.esa.int/archive/}} 
positions and position errors 
\citep{Gaia_Collaboration_2016,Gaia_Collaboration_2018b}
that we propagated
with 
{\tt topcat}\footnote{\href{http://www.star.bris.ac.uk/~mbt/topcat/}{http://www.star.bris.ac.uk/$\sim$mbt/topcat/}} 
\citep{taylor19} 
from the J2015.5 reference epoch to the J2017.948 \emph{Chandra} epoch 
(using the same zero radial velocity and radial velocity error as in the Celestia~2000, 
Hipparcos \& Tycho catalog)
and added 2MASS sources without the \emph{Gaia} DR2 best neighbor \citep{marrese19}; 
we excluded a 2MASS source that was resolved by \emph{Gaia} and \emph{Chandra}.
The new positions were also determined for isolated sources with an off-axis position larger than 5$^\prime$
using the PSF correlation position.
For crowded sources (PSF fraction < 87\%), the peak in a maximum-likelihood image 
reconstruction was used. 
The correlation and reconstruction were performed using a multi-ObsID event image 
and a composite PSF image.
These steps were repeated until converging to no offsets between the ObsIDs. 

\subsection{Astrometry}
\label{appendix:cxo_astrometry}

The final X-ray positions are mainly registered on \emph{Gaia} DR2 (with 45 \emph{Gaia} and 12 2MASS sources), 
that is to say, on the ICRS world coordinate system. The position of the X-ray counterpart of HOPS~383 
converted to the J2000 world coordinate system is 
05$^\mathrm{h}$35$^\mathrm{m}$29$\fs$789, -04$^\circ$59$^\prime$50$\farcs$432 
with a positional error of $0\farcs099$ that includes, added in quadrature, 
the first-ObsID reference offset uncertainty of $0\farcs029$ to 
the AE positional error of $0\farcs095$ 
\citep[Eqs.~(1)--(3) in][]{broos10}.

The radio position of JVLA-SE and NW in Figs.~\ref{fig:fov}b and c were determined 
from the pixel maximums in 
Fig.~1 of \cite{galvan-madrid15} converted to grayscale, 
assuming a J2000 world coordinate system, with positional errors
estimated using 
$FWHM/(2 \sqrt{2\ln{2}}\:S/N)$, where S/N is the signal-to-noise ratio.
For the Spitzer counterpart, we adopted a conservative position error 
of $1\arcsec$, corresponding to the matching radius used in \cite{megeath12}.

\subsection{Timing analysis}
\label{appendix:cxo_timing}

The source aperture with a PSF fraction of 96\% was used for the timing and spectral analysis of HOPS~383.
The background is negligible (Appendix~\ref{appendix:cxo_spectral}).
We estimated the count rates versus time ($\widehat{CR}$) 
from the arrival time
($t_i$) of the events ($N=28$) 
with a kernel estimator of the density \citep[$\widehat{f}$;][]{feigelson12b} 
with a constant bandwidth ($h$):
$\widehat{CR}(t, h)= N \widehat{f}(t, h)/ DTCOR$,
where $DTCOR=0.98693$ is the deadtime correction 
and $\widehat{f}(t, h)= (1/hN) \times \sum_{i=1}^N K\{(t-t_i)/h\}$,
where we used the Epanechnikov kernel, the inverted parabola defined 
as $K(y)=0.75\,(1-y^2)$ with $-1 \le y \le 1$.
We chose for $h$: the rule-of-thumb bandwidth, $h_\mathrm{rot}$, 
designed for unimodal distribution
\citep[Eq.~(6.9) in][and cyan curve in Fig~\ref{fig:lc}c]{feigelson12b}; 
and the optimal bandwidth, $h_\mathrm{cv}$, maximizing the cross-validation 
\citep[Eq.~(6.10) in][and green curve in Fig~\ref{fig:lc}]{feigelson12b}.
The latter is $\sim$3.6 larger than the former, 
producing a smoother light curve in particular at the end of the X-ray burst.

Then, we used an adaptive (bandwidth) kernel estimator  
\citep[Eqs.~(6.14)--(6.15) in][]{feigelson12b} 
with the recommend value of 0.5 for the sensitivity parameter $\alpha$. 
This was computed from the pilot densities previously obtained with $h_\mathrm{rot}$ (blue curve in Fig~\ref{fig:lc}c) 
and $h_\mathrm{cv}$ (red curve in Fig~\ref{fig:lc}c).

\subsection{Spectral analysis}
\label{appendix:cxo_spectral}

The source-plus-background spectrum was limited to the flare time interval 
ranging from the first to the last detected event
and the 0.5--9.9~keV energy range. 
An energy-dependent aperture correction was applied by AE
to the ancillary response file, which calibrates the extraction.
The background spectrum was extracted from the full exposure 
of the first observation
in the 0.1--10~keV energy range.

\begin{figure}[!t]
\centering
\begin{tabular}{@{}l@{}l@{}}
\raisebox{57mm}{\bf a} & \includegraphics[width=85mm,trim={0 0 0 0.32cm},clip]{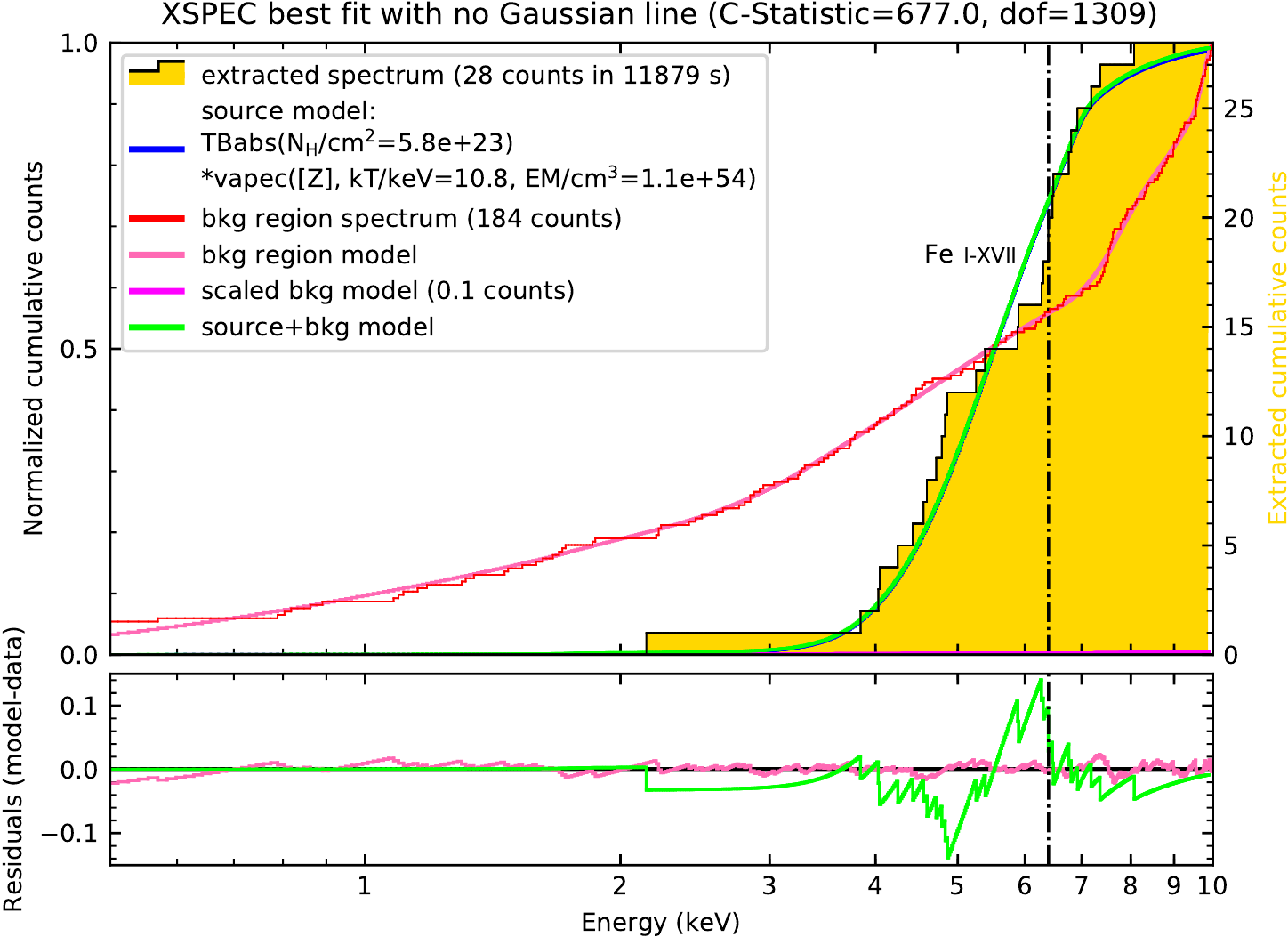} \\ 
\raisebox{57mm}{\bf b} & \includegraphics[width=85mm,trim={0 0 0 0.32cm},clip]{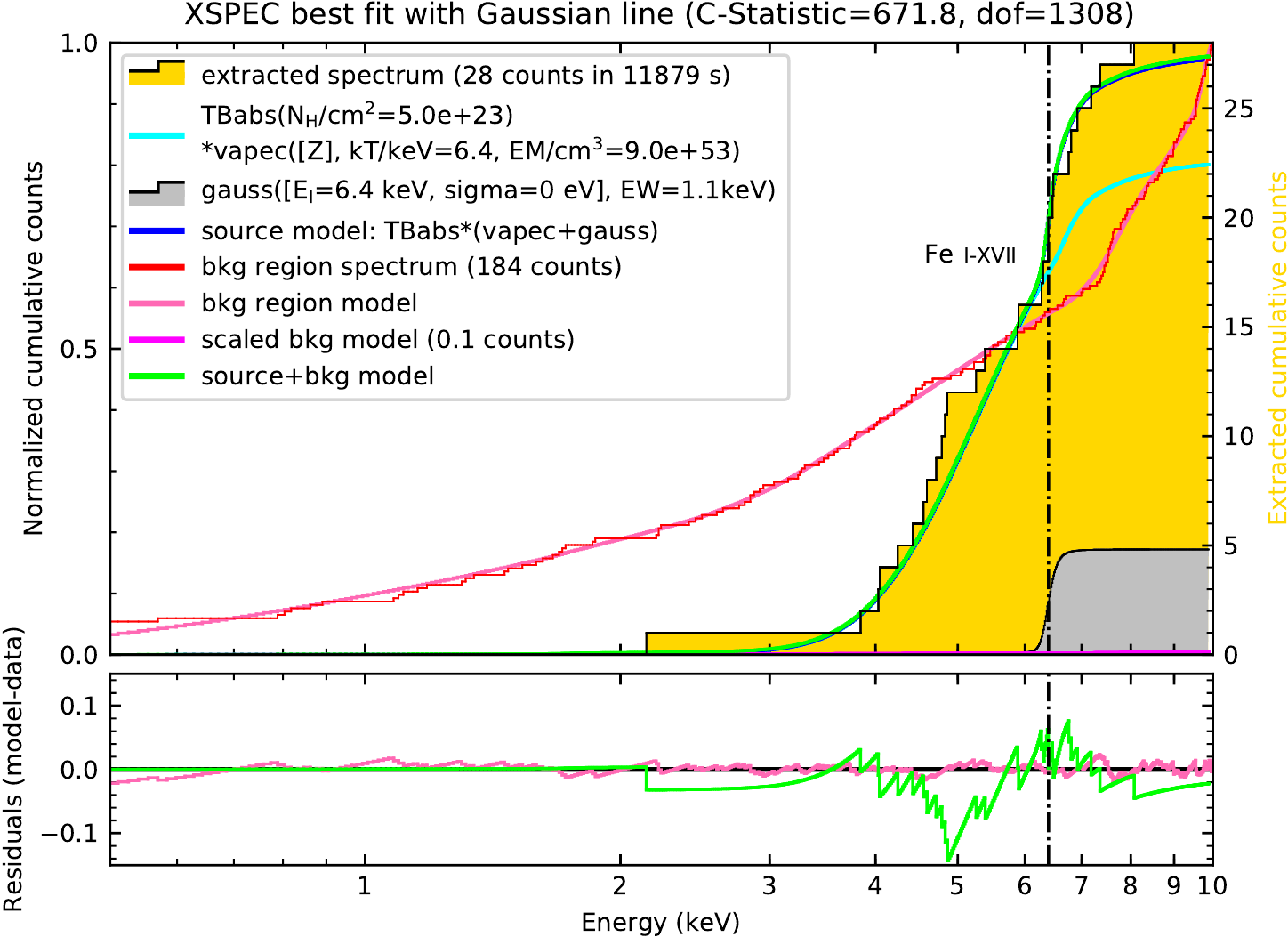}
\end{tabular}
\caption{X-ray spectrum of HOPS~383 and background determination. 
\emph{Panels~a and b}: \emph{top panels}: source-plus-background extracted cumulative spectrum (gold filled area) and {\tt XSPEC} best-fit model (green),
source model (blue),
background region cumulative spectrum (orange) and model (red),
and scaled background model (pink).
The vertical dashed-dotted line is the energy of the 
6.4~keV line arising from 
neutral or low-ionization iron.
\emph{Bottom panels}: model-minus-data residuals. 
\emph{Panel~b}:
Absorbed coronal thin emission (cyan) and 6.4~keV Gaussian emission line (silver filled area).
}
\label{appendix_fig:xspec_line}
\end{figure}

The simultaneous spectral fitting of the background and source-plus-background spectra was made with 
XSPEC\footnote{\href{https://heasarc.gsfc.nasa.gov/xanadu/xspec/}{https://heasarc.gsfc.nasa.gov/xanadu/xspec/}}
(version 12.10.1) 
using the C-statistic applied to unbinned data. 
The background cumulative
spectrum was modeled with the AE {\tt cplinear} model, 
a continuous piecewise-linear function with ten vertices \citep{broos10}.
The source spectrum was modeled with an interstellar medium absorption \citep{wilms00} 
and an isothermal collisional-radiative plasma \citep{smith01} 
with the typical coronal abundances of pre-main sequence stars \citep{guedel07}
plus an emission line with zero width at 6.4~keV  ({\tt TBabs*(vapec+gaussian)})
to which the background model, re-scaled to the source aperture, was added 
to fit the source-plus-background cumulative spectrum
\citep[data flow diagram in Fig.~10 of][]{broos10}.
The model parameters, 
namely the hydrogen column density, the plasma temperature and normalization, 
the line normalization, and the ten parameters of {\tt cplinear}, 
were first estimated by C-statistic minimization
(C-stat=677.0 using 1309 degrees of freedom; 
Fig.~\ref{appendix_fig:xspec_line}a).
The emission line was added to improve the modeling below $\sim$6.4~keV 
(C-stat=671.8 using 1308 degrees of freedom; 
Fig.~\ref{appendix_fig:xspec_line}b).

We fixed the {\tt cplinear} parameters since the background contribution to 
the source-plus-background spectrum (0.1~count) was negligible.
We performed a Markov chain Monte Carlo (MCMC) \citep{hogg18} to compute 
the probability density functions of the four physical parameters of this model 
using the Jeremy Sanders'~{\tt xspec\_emcee} 
program\footnote{\href{https://github.com/jeremysanders/xspec_emcee}{https://github.com/jeremysanders/xspec\_emcee}} 
that used a Python implementation of the MCMC ensemble sampler with 
affine invariance proposed 
by \cite{goodman10}, 
{\tt emcee}\footnote{\href{https://emcee.readthedocs.io/en/v2.2.1/}{https://emcee.readthedocs.io/en/v2.2.1/}} 
\citep[version 2.2.1;][]{foreman-mackey13}.
We customized {\tt xspec\_emcee} to output
subsidiary Markov chains for the acceptance fraction \citep{foreman-mackey13}
and the following physical values:
soft and hard fluxes; soft and hard absorption-corrected fluxes; and line equivalent width and counts.
We used uniform priors for the physical parameters, $N_\mathrm{w}=46$ walkers that were started clustered around the previous parameter 
estimates, and $N_\mathrm{iter}=10^6$ iterations.

\begin{figure}[!t]
\centering
\includegraphics[width=72mm,trim={0 0 0 0.5cm},clip]{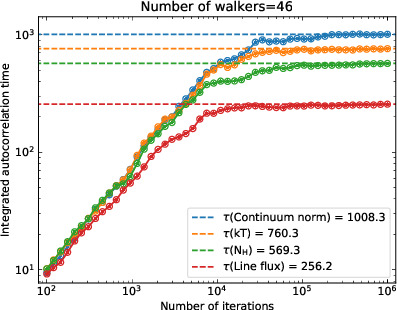}
\includegraphics[width=72mm,trim={0 0 0 0.5cm},clip]{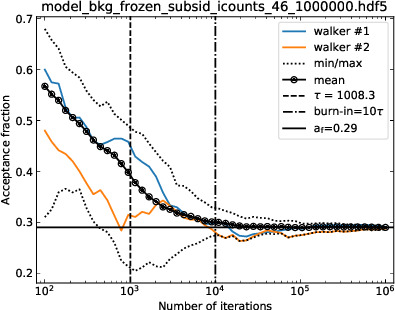}
\caption{Burn-in phase of the MCMC samples. 
\emph{Top panel}: estimator of the integrated autocorrelation time versus the number of iterations.
The dashed horizontal lines 
are the integrated autocorrelation time in an iteration-number unit for 
the model parameters
after 
$10^6$ iterations with 46 walkers; 
the burn-in phase is lower than ten times the maximum of these values. 
\emph{Bottom panel}: acceptance fraction versus the number of iterations. 
The mean of the walkers' acceptance fractions (dotted line) converges after the burn-in phase.
}
\label{appendix_fig:burn}
\end{figure}

To test for convergence of the Markov chains, we followed \cite{hogg18} by using the integrated autocorrelation time 
that we computed for each walker 
and then averaged these estimates\footnote{See Dan Foreman-Mackey's blog 
on autocorrelation time estimation (2017 October 16): 
\href{https://dfm.io/posts/autocorr/}{https://dfm.io/posts/autocorr/} .}
for each parameter, 
and retained the largest value, $\tau$ (top panel of Fig.~\ref{appendix_fig:burn}). 
The MCMC burn-in phase was supposed to last less than $N_\mathrm{burn{\mhyphen}in}=10\tau=10\,008$ 
iterations \citep{foreman-mackey13}, 
as this was illustrated by the mean of the walkers’ 
acceptance fractions converging to $\sim$0.3, which  
is well within the recommended range \citep{foreman-mackey13} of 0.2--0.5 
(bottom panel of Fig.~\ref{appendix_fig:burn}b). 


\begin{table*}[!t]
\caption{Markov chain Monte Carlo results on the flare time interval.}
\label{appendix_table:mcmc_results}
\centering
\begin{tabular}{lcccc}
\hline\hline\noalign{\vskip 1.2mm}
Parameter name & Symbol & \multicolumn{2}{c}{Value}      & Unit \\
\noalign{\vskip -2mm}
&             & \multicolumn{2}{c}{\hrulefill}      &\\
                    &             & Median & 90\% C.I. & \\
\hline\noalign{\vskip 1.2mm}
\multicolumn{5}{c}{Absorption}\\
\noalign{\vskip 1.2mm}
Hydrogen column density            & $N_\mathrm{H,X}$                 & 7.0    & 3.9--10.5   & $10^{23}$~cm$^{-2}$         \\
\noalign{\vskip 1.2mm}\hline\noalign{\vskip 1.2mm}
\multicolumn{5}{c}{Plasma emission}\\
\noalign{\vskip 1.2mm}
Temperature                        & $kT$                           & 4.1    & 1.3--17.8   & keV                         \\
                                   & $T$                            & 47.1   & 15.2--207.0 & MK                           \\
Emission measure\tablefootmark{\,(a)}             & $EM$                           & 2.1    & 0.5--32.5   & $10^{54}$~cm$^{-3}$         \\
Hard-band intrinsic luminosity\tablefootmark{\,(a)} & $L_\mathrm{X,intr}$(2-8~keV)   & 1.7    & 0.7--7.2    & $10^{31}$~erg~s$^{-1}$              \\
Soft-band intrinsic luminosity\tablefootmark{\,(a)} & $L_\mathrm{X,intr}$(0.5-2~keV) & 1.2    & 0.3--22.7   & $10^{31}$~erg~s$^{-1}$              \\
Full-band intrinsic luminosity\tablefootmark{\,(a)} & $L_\mathrm{X,intr}$(0.5-8~keV) & 3.0    & 0.9--29.9   & $10^{31}$~erg~s$^{-1}$              \\
\noalign{\vskip 1.2mm}\hline\noalign{\vskip 1.2mm}
\multicolumn{5}{c}{Neutral to low-ionization iron line emission}\\
\noalign{\vskip 1.2mm}
Line flux                          & Flux(Fe~6.4~keV)               & 6.1    & 1.7--14.1   & $10^{-6}$~ph~cm$^{-2}$~s$^{-1}$   \\
Equivalent width                   & EW(Fe~6.4~keV)                 & 1.1    & 0.2--3.1    & keV                         \\
Total count number                 & N(Fe~6.4~keV)                  & 5.1    & 1.4--10.8   & counts                           \\
\hline
\end{tabular}
\tablefoot{
\tablefoottext{a}{\!Assuming a distance of 420~pc \citep{safron15}.}
}
\end{table*}

\begin{figure}[!h]
\centering
\includegraphics[width=82mm]{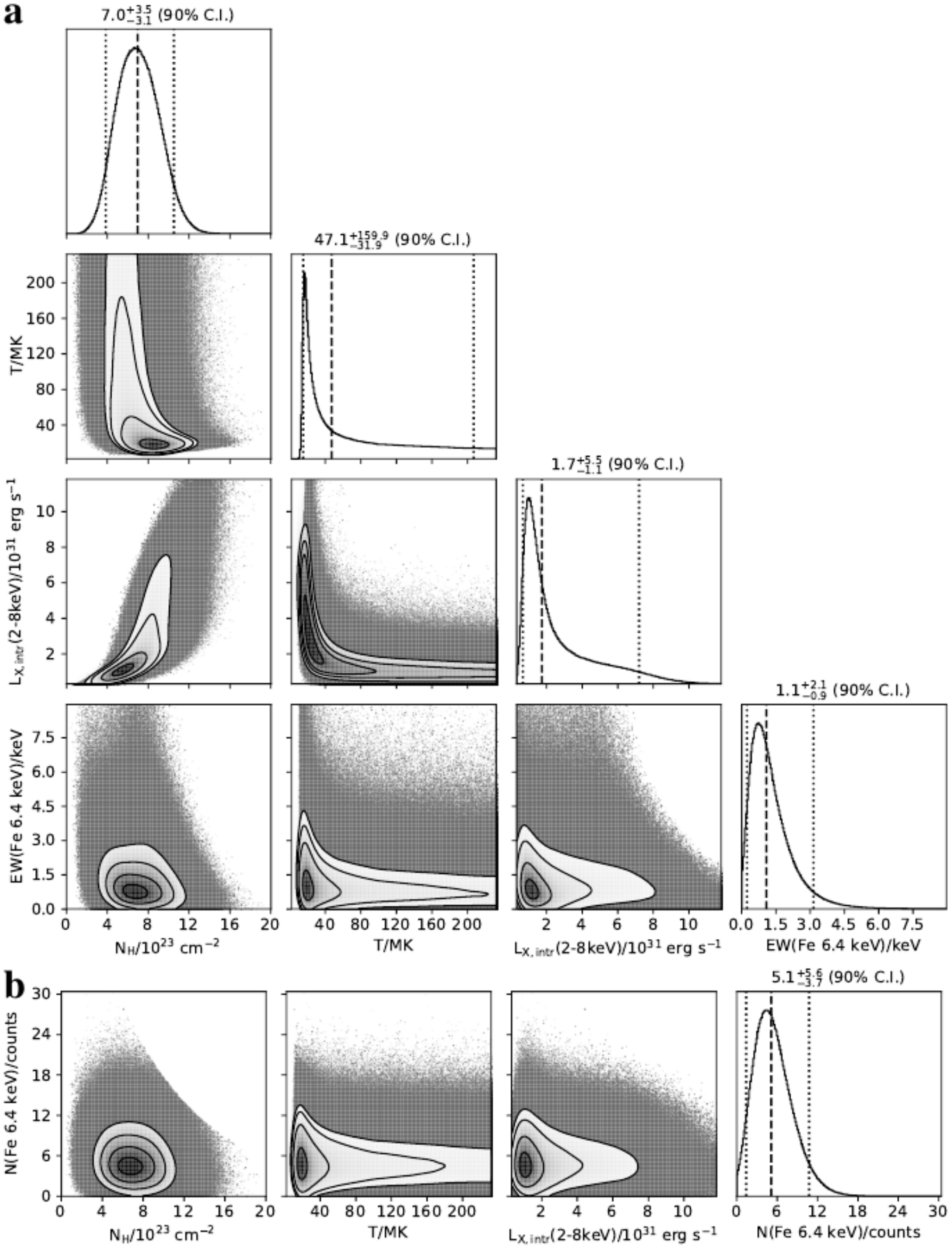}
\caption{ 
Posterior probability and covariance distributions 
of the flare-emission model parameters.  
\emph{Panel~a}: hydrogen column density, 
plasma temperature, 
X-ray intrinsic luminosity in the 2$-$8~keV range,
and equivalent width of the Fe
line at 6.4~keV.
The size of the MCMC samples is about $4.6\times10^7$ 
with about $4.5\times10^4$ independent MCMC samples.
The dashed and dotted vertical lines in the diagonal plots 
show 
the
median value and 
the
90\% confidence interval (C.I.) for each parameter. 
The contours in the other plots are 11.8, 39.3, 67.5, and 86.5\% C.I.
(corresponding to 0.5, 1, 1.5, and 2$\sigma$ levels for a 2D Gaussian) 
for each pair of parameters.
\emph{Panel~b}: alternative last row with count number from the emission line.
}
\label{appendix_fig:corner}
\end{figure}

\begin{figure}[!h]
\centering
\includegraphics[width=88mm,trim={0 0 0 0.33cm},clip]{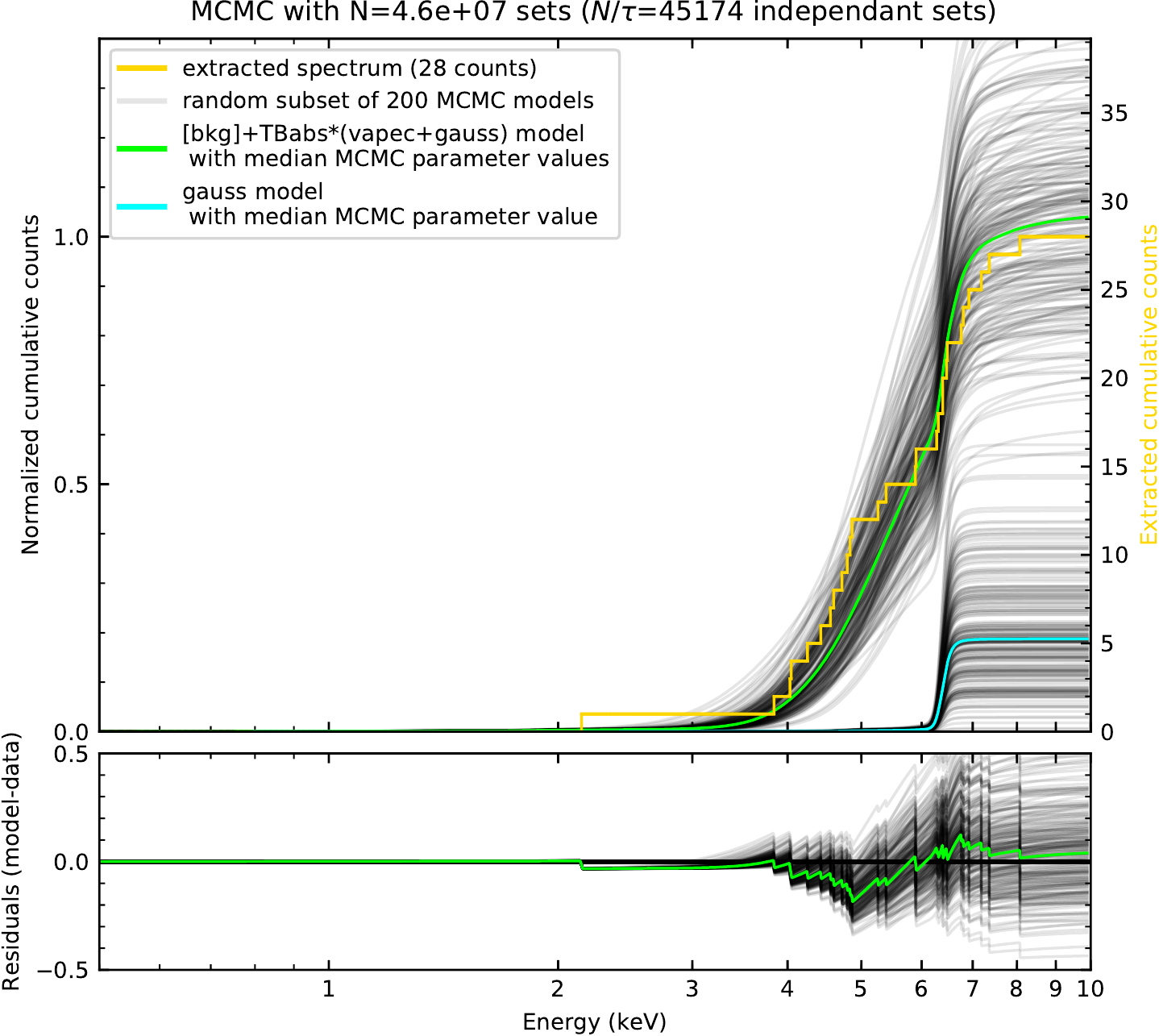}
\caption{X-ray spectrum of HOPS~383 and Markov chain Monte Carlo (MCMC) models.
\emph{Top panel}: source-plus-background extracted 
cumulative spectrum (gold) and median MCMC model (green),
6.4~keV Gaussian emission line with median MCMC model (cyan), 
and a random subset of 200 MCMC models (gray).
\emph{Bottom panel}: model-minus-data residuals.
}
\label{appendix_fig:spectrum}
\end{figure}

This MCMC burn-in phase was cut-off before producing the corner plots (Fig.~\ref{appendix_fig:corner}a)
from which the parameter median values 
and 90\% confidence interval\footnote{We followed 
\cite{hogg18}, who are in favor of reporting median and quantiles, 
which are based on integrals,
and against mode, which is not.}
were determined
(Table~\ref{appendix_table:mcmc_results}).
For comparison purposes, Fig.~\ref{appendix_fig:spectrum} shows 
the source-plus-background extracted 
cumulative counts spectrum (gold), the model for the parameter median values (green), and
a random subset of 200 MCMC models (gray).

In this MCMC, the number of effectively independent data points is
$N=(N_\mathrm{iter}-N_\mathrm{burn{\mhyphen}in}) \times N_\mathrm{w} \approx 4.6\times 10^7$
and the relative accuracy is $100\sqrt{\tau/N}=0.5\%$.
The probability of the emission line was computed 
as the probability to have at least 0.5~count from this line
(Fig.~\ref{appendix_fig:corner}b).

\subsection{Quiescent luminosity}
\label{appendix:quiescent}

We used the CIAO tool {\tt aprates} to compute the source count rate in the 2--8~keV energy band 
during our observations and excluding the flare time interval, which led to a live time of $\sim$72,000~s. 
The source-plus-background aperture centered on the source position had an area of $\sim$$2.4\protect{\arcsec}^2$ and 
corresponding to a PSF fraction of 90\% at 1.49~keV. 
The background aperture was a $\sim$$2\farcs5$--$\sim$30\arcsec\ annulus 
centered on the source position where the neighbor source region was excluded
with an area of $\sim$$2700\protect{\arcsec}^2$, corresponding to a PSF fraction of 0.9\% at 1.49~keV. 
The total counts in source-plus-background and background apertures were 1 (at 3.0~keV) and 304, respectively.
Inside the source-plus-background aperture, we estimated a quiescent source count rate of 0.011~count~ks$^{-1}$ 
(0--0.057~count~ks$^{-1}$, 90\% C.I.). 

From a {\tt tbabs*vapec} model with $N_\mathrm{H,X}=7.0\times10^{23}$~cm$^{-2}$
and a typical plasma quiescent temperature ($kT=1$~keV), we determined, using XSPEC, 
that the aperture corrected, observed flux and absorption-corrected luminosity 
were lower than $2.0\times10^{-15}$~erg~cm$^{-2}$~s$^{-1}$ and $2.0\times10^{30}$~erg~s$^{-1}$ 
in the 2--8~keV energy band, respectively.


\section{SOAR Spartan data} 
\label{appendix:spartan}

\subsection{Observations}
\label{appendix:soar_log}

The Spartan observations are listed in Table~\ref{appendix_table:soar_log}. 

\begin{table*}[!h]
\caption{Log of SOAR observations with Spartan 
(proposal ID 2016B-0930, PI: N.G.; observers: K.H.\ and D.P.).}
\label{appendix_table:soar_log}
\centering
\begin{tabular}{lccccccc}
\hline\hline\noalign{\vskip 1.2mm} 
Filter & Wavelength & Width & Start Time &  DIT\tablefootmark{\,(a)} & NDIT\tablefootmark{\,(b)} & Exposure\tablefootmark{\,(c)} & FWHM\tablefootmark{\,(d)}\\
            & (microns)        & (microns)    & (UT)     &     (s)        &         & (s)              &   (arcsec)\\
\hline\noalign{\vskip 1.2mm}
$K$             & 2.148 & 0.307 & 2017-12-14T01:01:22  & 179.98 & 37 & 6659.26 & 1.0\\
H$_\mathrm{2}$  & 2.116 & 0.031 & 2017-12-14T03:47:50  & 240.09 & 11 & 2640.99 & 1.0\\
\hline
\end{tabular}
\tablefoot{
We also obtained standard infrared calibration data: dark exposures and dome flat exposures with the lamp on and off to account 
for the thermal emission from the telescope.
\tablefoottext{a}{\!Detector integration time.}
\tablefoottext{b}{\!Number of frame exposures with the target.}
\tablefoottext{c}{\!Total exposure of the stacked dithered frames.}
\tablefoottext{d}{\!Image quality of the stacked dithered frames.}
}
\end{table*}

\subsection{Data reduction}
\label{appendix:soar_reduction}

An end-to-end pipeline, THELI\footnote{\href{https://www.astro.uni-bonn.de/theli/}{https://www.astro.uni-bonn.de/theli/}}
\citep[version 1.9.5;][]{schirmer13},
was used for the reduction of the calibration and observation data 
(Table~\ref{appendix_table:soar_log})
obtained with det3, 
including astrometry with {\tt SExtractor} \citep{bertin96} and {\tt SWarp} \citep{bertin02}.
We wrote a custom IDL program to automatically detect and correct, in the dark and flat-field corrected frames,
any residual bulk images that were produced by bright stars, close to saturation in the raw frames.
In each raw frame, this program identified any saturated regions centered on bright stars 
and looked at these detector positions in the subsequent calibrated frames for any excess above the sky level 
by fitting a constant level or a parabola (based on an F-test). 
These excesses were then subtracted or masked when too close to a source.
These calibrated and corrected images were registered using the 2MASS catalog and stacked using median values
and a pixel scale of $0\farcs5$.

In the H$_2$ narrow-band filter image,
the shocked molecular hydrogen emission 
has a higher 
S/N
compared to the $K$-band image. 
We produce for validation a pure H$_2$ emission image by subtracting the $K$-band image scaled by a factor 
that we obtain by minimizing the residuals on stars in the final image \citep{navarete15}.

\section{Density model}
\label{appendix:model}

We started from the protostellar model of HOPS~383, 
which is composed of an accretion disk in an infalling envelope with bipolar cavities 
\citep{furlan16}.
The fixed parameters in this model grid  
\citep[Table~3 of][]{furlan16}
that was computed with the
Monte Carlo radiative transfer code, {\tt ttsre}\footnote{\href{https://gemelli.colorado.edu/~bwhitney/codes/codes.html}{https://gemelli.colorado.edu/$\sim$bwhitney/codes/codes.html}}
\citep[2008 version;][]{whitney03} were:
   \begin{itemize}
   \item the stellar mass ($M_\star=0.5$~$M_\odot$), 
   \item the stellar effective temperature ($T_\star=4000$~K), 
   \item the disk mass ($M_\mathrm{disk}=0.05$~$M_\odot$),
   \item the magnetospheric truncation radius of the gas disk ($R_\mathrm{trunc}=3$~$R_\star$),
   \item the dust-disk inner radius 
(set to the dust sublimation radius, 
$R_\mathrm{sub}/R_\star=(T_\mathrm{sub}/T_\star)^{-2.1}=6.8$ 
where $T_\mathrm{sub}=1600$~K is the dust sublimation temperature; 
\citealt{whitney03}),
   \item the scale height of the disk at the dust sublimation radius 
(set to the hydrostatic equilibrium solution; Eq.~(35) of \citealt{dalessio98}),
   \item the mean molecular mass per hydrogen nucleus ($\mu=2.3$),
   \item the radial ($A=2.25$) and vertical ($B=1.25$) exponent in disk density law,
   \item the fractional area of the hot spots on the star ($f_\mathrm{spot}=0.01$),
   \item the envelope outer radius ($R_\mathrm{env}=10\,000$~au),
   \item the centrifugal radius of the envelope ($R_\mathrm{c}=R_\mathrm{disk}$),
   \item the exponent of the cavity polynomial shape ($b_\mathrm{cav}=1.5$),
and   \item the vertical offset of the cavity wall ($z_\mathrm{cav}=0$~au).
   \end{itemize}

The best-fit parameters 
for HOPS~383
(Table~1 of \citealt{furlan16})
were:
   \begin{itemize}
   \item the stellar radius ($R_\star=6.61$~$R_\odot$),
   \item the disk outer radius ($R_\mathrm{disk}=5$~au; we
    note that the opacity was dominated by the envelope in these SED models, 
    which did not directly constrain the size of the disk; \citealt{furlan16}),
   \item the envelope (gas and dust) density at 1000~au ($\rho_{1000}=1.78\times10^{-17}$~g~cm$^{-3}$)
    and, 
    as $\rho_{1000} \propto \dot{M}_\mathrm{env}$, 
    the envelope infall rate 
    ($\dot{M}_\mathrm{env}=7.48\times10^{-5}$~$M_\odot$~yr$^{-1}$),
   \item the cavity opening angle ($\theta=45^\circ$),
   \item the stellar luminosity ($L_\star=10$~$L_\odot$),
   \item the total (star and accretion) luminosity 
    ($L_\mathrm{tot}=s\times30.2$~$L_\odot=55.49$~$L_\odot$,
    where $s=1.84$ is the luminosity scaling factor to match the observed SED 
    flux, Eq.~(3) of \citealt{furlan16}),
   \item the inclination angle ($i=69.5^\circ$; corresponding in the model grid to $\cos{i}=0.35$ with 
    a sampling step of 0.1),
and   \item the foreground cloud extinction ($A_\mathrm{V, cloud}=14$~mag; 
    corresponding to 
    $N_\mathrm{H, cloud}=14\times10^{21}$~cm$^{-2}$,
    as $N_\mathrm{H, cloud}/A_\mathrm{V, cloud}=1.0\times10^{21}$~cm$^{-2}$~mag$^{-1}$ 
    in \citealt{furlan16}).
   \end{itemize}

To be consistent with the 
luminosity scaling factor ($s$), we introduced $s^\prime$, a scaling factor for 
the disk-to-star accretion rate ($\dot{M}_\mathrm{disk}$). 
Indeed, 
the above values of
$R_\star$ (6.61~$R_\odot$), 
$L_\star$ (10~$L_\odot$), 
and $L_\mathrm{tot}=L_\star+L_\mathrm{acc}$ 
(30.2~$L_\odot$) 
correspond 
in the model grid
to $\dot{M}_\mathrm{disk}=1.14\times10^{-5}$~$M_\odot$~yr$^{-1}$, since $L_\mathrm{acc} \propto \dot{M}_\mathrm{disk}$, 
we set $s^\prime=2.25$ and hence $\dot{M}_\mathrm{disk}=2.57\times10^{-5}$~$M_\odot$~yr$^{-1}$.

The envelope density was given by Eqs.~(1) and (2) of \cite{whitney03}. 
We note that the bipolar cavities in \cite{furlan16} are free of dust (and gas); 
for consistency, we filled 
them in
with gas, using
a constant 
density 
\citep[$n_\mathrm{H_2}=2\times10^4$~cm$^{-3}$;][]{whitney03}.

The disk density was given by Eq.~(3) of \cite{whitney03} 
and normalized to $M_\mathrm{disk}$ from $R_\mathrm{trunc}$ to $R_\mathrm{disk}$ 
using a spherical outer boundary (Fig.~\ref{fig:nh}b this work and Fig.~2b of \citealt{whitney03}).
We added accretion funnels at $R_\mathrm{trunc}$, which were dust-free as located inside the dust sublimation radius, 
assuming for simplicity a dipole geometry for the magnetic field \citep{hartmann94}, 
and using $f_\mathrm{spot}$ to compute a latitude band between 
the smallest and largest magnetic loops at the stellar surface. 
The accretion-funnel density was given by Eq.~(9) of \cite{hartmann94}.

The values of $N_\mathrm{H}$ along the line-of-sight were computed 
by numerically integrating these density equations. 
In this model grid, the gas-to-dust ratio is $R=100$ and 
the 
protostellar
dust opacity in the visible is $\kappa_\mathrm{ext, V}=5.3\times10^4$~cm$^2$~g$^{-1}$ 
(Fig.~6 of \citealt{ormel11}, (ic-sil,gra) at 0.3~Myr), 
leading 
to:
$N_\mathrm{H}/A_\mathrm{V}=\ln{10}\times(1+R)/(2.5\,\mu\, m_\mathrm{H}\,\kappa_\mathrm{ext, V})=4.6\times10^{20}$~cm$^{-2}$~mag$^{-1}$
where $m_\mathrm{H}$ is the hydrogen mass. 
Therefore, the observed $N_\mathrm{H, X}$ 
corresponds to $A_\mathrm{V} \sim 1500$~mag.

\end{appendix}


\begin{thebibliography}{72}
\expandafter\ifx\csname natexlab\endcsname\relax\def\natexlab#1{#1}\fi

\bibitem[{{Andr{\'e}} {et~al.}(2008){Andr{\'e}}, {Minier}, {Gallais},
  {Rev{\'e}ret}, {Le Pennec}, {Rodriguez}, {Boulade}, {Doumayrou}, {Dubreuil},
  {Lortholary}, {Martignac}, {Talvard}, {De Breuck}, {Hamon}, {Schneider},
  {Bontemps}, {Lagage}, {Pantin}, {Roussel}, {Miller}, {Purcell}, {Hill}, \&
  {Stutzki}}]{andre08}
{Andr{\'e}}, P., {Minier}, V., {Gallais}, P., {et~al.} 2008, \aap, 490, L27

\bibitem[{{Andr\'e} \& {Montmerle}(1994)}]{andre94}
{Andr\'e}, P. \& {Montmerle}, T. 1994, \apj, 420, 837

\bibitem[{{Andr\'e} {et~al.}(1993){Andr\'e}, {Ward-Thompson}, \&
  {Barsony}}]{andre93}
{Andr\'e}, P., {Ward-Thompson}, D., \& {Barsony}, M. 1993, \apj, 406, 122

\bibitem[{{Andr\'e} {et~al.}(2000){Andr\'e}, {Ward-Thompson}, \&
  {Barsony}}]{andre00}
{Andr\'e}, P., {Ward-Thompson}, D., \& {Barsony}, M. 2000, in Protostars \&
  Planets IV, ed. V.~{Mannings}, A.~P. {Boss}, \& S.~S. {Russell} (Tucson:
  University of Arizona Press), 59--95

\bibitem[{{Barret} {et~al.}(2018){Barret}, {Lam Trong}, {den Herder}, {Piro},
  {Cappi}, {Houvelin}, {Kelley}, {Mas-Hesse}, {Mitsuda}, {Paltani}, {Rauw},
  {Rozanska}, {Wilms}, {Bandler}, {Barbera}, {Barcons}, {Bozzo}, {Ceballos},
  {Charles}, {Costantini}, {Decourchelle}, {den Hartog}, {Duband}, {Duval},
  {Fiore}, {Gatti}, {Goldwurm}, {Jackson}, {Jonker}, {Kilbourne}, {Macculi},
  {Mendez}, {Molendi}, {Orleanski}, {Pajot}, {Pointecouteau}, {Porter},
  {Pratt}, {Pr{\^e}le}, {Ravera}, {Sato}, {Schaye}, {Shinozaki}, {Thibert},
  {Valenziano}, {Valette}, {Vink}, {Webb}, {Wise}, {Yamasaki}, {Douchin},
  {Mesnager}, {Pontet}, {Pradines}, {Branduardi-Raymont}, {Bulbul}, {Dadina},
  {Ettori}, {Finoguenov}, {Fukazawa}, {Janiuk}, {Kaastra}, {Mazzotta},
  {Miller}, {Miniutti}, {Naze}, {Nicastro}, {Scioritino}, {Simonescu},
  {Torrejon}, {Frezouls}, {Geoffray}, {Peille}, {Aicardi}, {Andr{\'e}},
  {Daniel}, {Cl{\'e}net}, {Etcheverry}, {Gloaguen}, {Hervet}, {Jolly}, {Ledot},
  {Paillet}, {Schmisser}, {Vella}, {Damery}, {Boyce}, {Dipirro}, {Lotti},
  {Schwander}, {Smith}, {Van Leeuwen}, {van Weers}, {Clerc}, {Cobo}, {Dauser},
  {Kirsch}, {Cucchetti}, {Eckart}, {Ferrando}, \& {Natalucci}}]{barret18}
{Barret}, D., {Lam Trong}, T., {den Herder}, J.-W., {et~al.} 2018, in Space Telescopes and Instrumentation 2018: Ultraviolet
  to Gamma Ray, \procspie, 10699, 106991G

\bibitem[{{Bertin} \& {Arnouts}(1996)}]{bertin96}
{Bertin}, E. \& {Arnouts}, S. 1996, \aaps, 117, 393

\bibitem[{{Bertin} {et~al.}(2002){Bertin}, {Mellier}, {Radovich}, {Missonnier},
  {Didelon}, \& {Morin}}]{bertin02}
{Bertin}, E., {Mellier}, Y., {Radovich}, M., {et~al.} 2002, in Astronomical Data Analysis Software and Systems XI, 
ed. D.~A. {Bohlender}, D.~{Durand}, \& T.~H. {Handley},
  ASP Conf.\ Ser., 281, 228

\bibitem[{{Bontemps} {et~al.}(1996){Bontemps}, {Andr\'e}, {Terebey}, \&
  {Cabrit}}]{bontemps96}
{Bontemps}, S., {Andr\'e}, P., {Terebey}, S., \& {Cabrit}, S. 1996, \aap, 311,
  858

\bibitem[{{Broos} {et~al.}(2010){Broos}, {Townsley}, {Feigelson}, {Getman},
  {Bauer}, \& {Garmire}}]{broos10}
{Broos}, P.~S., {Townsley}, L.~K., {Feigelson}, E.~D., {et~al.} 2010, \apj,
  714, 1582

\bibitem[{{Ceccarelli} {et~al.}(2014){Ceccarelli}, {Dominik},
  {L{\'o}pez-Sepulcre}, {Kama}, {Padovani}, {Caux}, \&
  {Caselli}}]{ceccarelli14}
{Ceccarelli}, C., {Dominik}, C., {L{\'o}pez-Sepulcre}, A., {et~al.} 2014,
  \apjl, 790, L1

\bibitem[{{Czesla} \& {Schmitt}(2010)}]{czesla10}
{Czesla}, S. \& {Schmitt}, J.~H.~M.~M. 2010, \aap, 520, A38

\bibitem[{{D'Alessio} {et~al.}(1998){D'Alessio}, {Canto}, {Calvet}, \&
  {Lizano}}]{dalessio98}
{D'Alessio}, P., {Canto}, J., {Calvet}, N., \& {Lizano}, S. 1998, \apj, 500,
  411

\bibitem[{{Davis} {et~al.}(2012){Davis}, {Bautz}, {Dewey}, {Heilmann}, {Houck},
  {Huenemoerder}, {Marshall}, {Nowak}, {Schattenburg}, {Schulz}, \&
  {Smith}}]{davis12}
{Davis}, J.~E., {Bautz}, M.~W., {Dewey}, D., {et~al.} 2012, in Space Telescopes and Instrumentation 2012: Ultraviolet to Gamma Ray, 
\procspie, 8443, 84431A

\bibitem[{{Drake} {et~al.}(2008){Drake}, {Ercolano}, \& {Swartz}}]{drake08}
{Drake}, J.~J., {Ercolano}, B., \& {Swartz}, D.~A. 2008, \apj, 678, 385

\bibitem[{{Dunham} {et~al.}(2014){Dunham}, {Stutz}, {Allen}, {Evans},
  {Fischer}, {Megeath}, {Myers}, {Offner}, {Poteet}, {Tobin}, \&
  {Vorobyov}}]{dunham14}
{Dunham}, M.~M., {Stutz}, A.~M., {Allen}, L.~E., {et~al.} 2014, in Protostars
  \& Planets VI, ed. H.~{Beuther}, R.~{Klessen}, C.~{Dullemond}, \&
  T.~{Henning} (Univ.\ of Arizona Space Sci.\ Series), 195--218

\bibitem[{{Dzib} {et~al.}(2013){Dzib}, {Loinard}, {Mioduszewski},
  {Rodr{\'{\i}}guez}, {Ortiz-Le{\'o}n}, {Pech}, {Rivera}, {Torres}, {Boden},
  {Hartmann}, {Evans}, {Brice{\~n}o}, \& {Tobin}}]{dzib13}
{Dzib}, S.~A., {Loinard}, L., {Mioduszewski}, A.~J., {et~al.} 2013, \apj, 775,
  63

\bibitem[{Feddersen {et~al.}(2020)Feddersen, Arce, Kong, Suri, Álvaro
  Sánchez-Monge, Ossenkopf-Okada, Dunham, Nakamura, Shimajiri, \&
  Bally}]{feddersen20}
Feddersen, J.~R., Arce, H.~G., Kong, S., {et~al.} 2020, \apj, accepted
  [arXiv:2004.03504]

\bibitem[{{Feigelson} {et~al.}(2002){Feigelson}, {Garmire}, \&
  {Pravdo}}]{feigelson02}
{Feigelson}, E.~D., {Garmire}, G.~P., \& {Pravdo}, S.~H. 2002, \apj, 572, 335

\bibitem[{{Feigelson} {et~al.}(2005){Feigelson}, {Getman}, {Townsley},
  {Garmire}, {Preibisch}, {Grosso}, {Montmerle}, {Muench}, \&
  {McCaughrean}}]{feigelson05}
{Feigelson}, E.~D., {Getman}, K., {Townsley}, L., {et~al.} 2005, \apjs, 160,
  379

\bibitem[{{Feigelson} \& {Jogesh Babu}(2012)}]{feigelson12b}
{Feigelson}, E.~D. \& {Jogesh Babu}, G. 2012, {Modern Statistical Methods for
  Astronomy} (Cambridge University Press)

\bibitem[{{Feigelson} \& {Montmerle}(1999)}]{feigelson99}
{Feigelson}, E.~D. \& {Montmerle}, T. 1999, \araa, 37, 363

\bibitem[{{Fischer} \& {Hillenbrand}(2017)}]{fischer17c}
{Fischer}, W.~J. \& {Hillenbrand}, L. 2017, ATel, 9969

\bibitem[{{Foreman-Mackey} {et~al.}(2013){Foreman-Mackey}, {Hogg}, {Lang}, \&
  {Goodman}}]{foreman-mackey13}
{Foreman-Mackey}, D., {Hogg}, D.~W., {Lang}, D., \& {Goodman}, J. 2013, \pasp,
  125, 306

\bibitem[{{Fruscione} {et~al.}(2006){Fruscione}, {McDowell}, {Allen},
  {Brickhouse}, {Burke}, {Davis}, {Durham}, {Elvis}, {Galle}, {Harris},
  {Huenemoerder}, {Houck}, {Ishibashi}, {Karovska}, {Nicastro}, {Noble},
  {Nowak}, {Primini}, {Siemiginowska}, {Smith}, \& {Wise}}]{fruscione06}
{Fruscione}, A., {McDowell}, J.~C., {Allen}, G.~E., {et~al.} 2006, 
in Observatory Operations: Strategies, Processes, and Systems,
  \procspie, 6270, 62701V

\bibitem[{{Furlan} {et~al.}(2016){Furlan}, {Fischer}, {Ali}, {Stutz}, {Stanke},
  {Tobin}, {Megeath}, {Osorio}, {Hartmann}, {Calvet}, {Poteet}, {Booker},
  {Manoj}, {Watson}, \& {Allen}}]{furlan16}
{Furlan}, E., {Fischer}, W.~J., {Ali}, B., {et~al.} 2016, \apjs, 224, 5

\bibitem[{{Gaia Collaboration} {et~al.}(2018){Gaia Collaboration}, {Brown},
  {Vallenari}, {Prusti}, {de Bruijne}, {Babusiaux}, {Bailer-Jones}, {Biermann},
  {Evans}, {Eyer}, {Jansen}, {Jordi}, {Klioner}, {Lammers}, {Lindegren},
  {Luri}, {Mignard}, {Panem}, {Pourbaix}, {Randich}, {Sartoretti}, {Siddiqui},
  {Soubiran}, {van Leeuwen}, {Walton}, {Arenou}, {Bastian}, {Cropper},
  {Drimmel}, {Katz}, {Lattanzi}, {Bakker}, {Cacciari}, {Casta{\~n}eda},
  {Chaoul}, {Cheek}, {De Angeli}, {Fabricius}, {Guerra}, {Holl}, {Masana},
  {Messineo}, {Mowlavi}, {Nienartowicz}, {Panuzzo}, {Portell}, {Riello},
  {Seabroke}, {Tanga}, {Th{\'e}venin}, {Gracia-Abril}, {Comoretto},
  {Garcia-Reinaldos}, {Teyssier}, {Altmann}, {Andrae}, {Audard},
  {Bellas-Velidis}, {Benson}, {Berthier}, {Blomme}, {Burgess}, {Busso},
  {Carry}, {Cellino}, {Clementini}, {Clotet}, {Creevey}, {Davidson}, {De
  Ridder}, {Delchambre}, {Dell'Oro}, {Ducourant},
  {Fern{\'a}ndez-Hern{\'a}ndez}, {Fouesneau}, {Fr{\'e}mat}, {Galluccio},
  {Garc{\'\i}a-Torres}, {Gonz{\'a}lez-N{\'u}{\~n}ez}, {Gonz{\'a}lez-Vidal},
  {Gosset}, {Guy}, {Halbwachs}, {Hambly}, {Harrison}, {Hern{\'a}ndez},
  {Hestroffer}, {Hodgkin}, {Hutton}, {Jasniewicz}, {Jean-Antoine-Piccolo},
  {Jordan}, {Korn}, {Krone-Martins}, {Lanzafame}, {Lebzelter}, {L{\"o}ffler},
  {Manteiga}, {Marrese}, {Mart{\'\i}n-Fleitas}, {Moitinho}, {Mora}, {Muinonen},
  {Osinde}, {Pancino}, {Pauwels}, {Petit}, {Recio-Blanco}, {Richards},
  {Rimoldini}, {Robin}, {Sarro}, {Siopis}, {Smith}, {Sozzetti}, {S{\"u}veges},
  {Torra}, {van Reeven}, {Abbas}, {Abreu Aramburu}, {Accart}, {Aerts},
  {Altavilla}, {{\'A}lvarez}, {Alvarez}, {Alves}, {Anderson}, {Andrei},
  {Anglada Varela}, {Antiche}, {Antoja}, {Arcay}, {Astraatmadja}, {Bach},
  {Baker}, {Balaguer-N{\'u}{\~n}ez}, {Balm}, {Barache}, {Barata}, {Barbato},
  {Barblan}, {Barklem}, {Barrado}, {Barros}, {Barstow}, {Bartholom{\'e}
  Mu{\~n}oz}, {Bassilana}, {Becciani}, {Bellazzini}, {Berihuete}, {Bertone},
  {Bianchi}, {Bienaym{\'e}}, {Blanco-Cuaresma}, {Boch}, {Boeche}, {Bombrun},
  {Borrachero}, {Bossini}, {Bouquillon}, {Bourda}, {Bragaglia}, {Bramante},
  {Breddels}, {Bressan}, {Brouillet}, {Br{\"u}semeister}, {Brugaletta},
  {Bucciarelli}, {Burlacu}, {Busonero}, {Butkevich}, {Buzzi}, {Caffau},
  {Cancelliere}, {Cannizzaro}, {Cantat-Gaudin}, {Carballo}, {Carlucci},
  {Carrasco}, {Casamiquela}, {Castellani}, {Castro-Ginard}, {Charlot},
  {Chemin}, {Chiavassa}, {Cocozza}, {Costigan}, {Cowell}, {Crifo}, {Crosta},
  {Crowley}, {Cuypers}, {Dafonte}, {Damerdji}, {Dapergolas}, {David}, {David},
  {de Laverny}, {De Luise}, {De March}, {de Martino}, {de Souza}, {de Torres},
  {Debosscher}, {del Pozo}, {Delbo}, {Delgado}, {Delgado}, {Di Matteo},
  {Diakite}, {Diener}, {Distefano}, {Dolding}, {Drazinos}, {Dur{\'a}n},
  {Edvardsson}, {Enke}, {Eriksson}, {Esquej}, {Eynard Bontemps}, {Fabre},
  {Fabrizio}, {Faigler}, {Falc{\~a}o}, {Farr{\`a}s Casas}, {Federici},
  {Fedorets}, {Fernique}, {Figueras}, {Filippi}, {Findeisen}, {Fonti},
  {Fraile}, {Fraser}, {Fr{\'e}zouls}, {Gai}, {Galleti}, {Garabato},
  {Garc{\'\i}a-Sedano}, {Garofalo}, {Garralda}, {Gavel}, {Gavras}, {Gerssen},
  {Geyer}, {Giacobbe}, {Gilmore}, {Girona}, {Giuffrida}, {Glass}, {Gomes},
  {Granvik}, {Gueguen}, {Guerrier}, {Guiraud}, {Guti{\'e}rrez-S{\'a}nchez},
  {Haigron}, {Hatzidimitriou}, {Hauser}, {Haywood}, {Heiter}, {Helmi}, {Heu},
  {Hilger}, {Hobbs}, {Hofmann}, {Holland}, {Huckle}, {Hypki}, {Icardi},
  {Jan{\ss}en}, {Jevardat de Fombelle}, {Jonker}, {Juh{\'a}sz}, {Julbe},
  {Karampelas}, {Kewley}, {Klar}, {Kochoska}, {Kohley}, {Kolenberg},
  {Kontizas}, {Kontizas}, {Koposov}, {Kordopatis}, {Kostrzewa-Rutkowska},
  {Koubsky}, {Lambert}, {Lanza}, {Lasne}, {Lavigne}, {Le Fustec}, {Le
  Poncin-Lafitte}, {Lebreton}, {Leccia}, {Leclerc}, {Lecoeur-Taibi},
  {Lenhardt}, {Leroux}, {Liao}, {Licata}, {Lindstr{\o}m}, {Lister}, {Livanou},
  {Lobel}, {L{\'o}pez}, {Managau}, {Mann}, {Mantelet}, {Marchal}, {Marchant},
  {Marconi}, {Marinoni}, {Marschalk{\'o}}, {Marshall}, {Martino}, {Marton},
  {Mary}, {Massari}, {Matijevi{\v{c}}}, {Mazeh}, {McMillan}, {Messina},
  {Michalik}, {Millar}, {Molina}, {Molinaro}, {Moln{\'a}r}, {Montegriffo},
  {Mor}, {Morbidelli}, {Morel}, {Morris}, {Mulone}, {Muraveva}, {Musella},
  {Nelemans}, {Nicastro}, {Noval}, {O'Mullane}, {Ord{\'e}novic},
  {Ord{\'o}{\~n}ez-Blanco}, {Osborne}, {Pagani}, {Pagano}, {Pailler},
  {Palacin}, {Palaversa}, {Panahi}, {Pawlak}, {Piersimoni}, {Pineau}, {Plachy},
  {Plum}, {Poggio}, {Poujoulet}, {Pr{\v{s}}a}, {Pulone}, {Racero}, {Ragaini},
  {Rambaux}, {Ramos-Lerate}, {Regibo}, {Reyl{\'e}}, {Riclet}, {Ripepi}, {Riva},
  {Rivard}, {Rixon}, {Roegiers}, {Roelens}, {Romero-G{\'o}mez}, {Rowell},
  {Royer}, {Ruiz-Dern}, {Sadowski}, {Sagrist{\`a} Sell{\'e}s}, {Sahlmann},
  {Salgado}, {Salguero}, {Sanna}, {Santana-Ros}, {Sarasso}, {Savietto},
  {Schultheis}, {Sciacca}, {Segol}, {Segovia}, {S{\'e}gransan}, {Shih},
  {Siltala}, {Silva}, {Smart}, {Smith}, {Solano}, {Solitro}, {Sordo}, {Soria
  Nieto}, {Souchay}, {Spagna}, {Spoto}, {Stampa}, {Steele},
  {Steidelm{\"u}ller}, {Stephenson}, {Stoev}, {Suess}, {Surdej}, {Szabados},
  {Szegedi-Elek}, {Tapiador}, {Taris}, {Tauran}, {Taylor}, {Teixeira},
  {Terrett}, {Teyssand ier}, {Thuillot}, {Titarenko}, {Torra Clotet}, {Turon},
  {Ulla}, {Utrilla}, {Uzzi}, {Vaillant}, {Valentini}, {Valette}, {van Elteren},
  {Van Hemelryck}, {van Leeuwen}, {Vaschetto}, {Vecchiato}, {Veljanoski},
  {Viala}, {Vicente}, {Vogt}, {von Essen}, {Voss}, {Votruba}, {Voutsinas},
  {Walmsley}, {Weiler}, {Wertz}, {Wevers}, {Wyrzykowski}, {Yoldas},
  {{\v{Z}}erjal}, {Ziaeepour}, {Zorec}, {Zschocke}, {Zucker}, {Zurbach}, \&
  {Zwitter}}]{Gaia_Collaboration_2018b}
{Gaia Collaboration (Brown, A.~G.~A., et~al.)} 2018, \aap,
  616, A1

\bibitem[{{Gaia Collaboration} {et~al.}(2016){Gaia Collaboration}, {Prusti},
  {de Bruijne}, {Brown}, {Vallenari}, {Babusiaux}, {Bailer-Jones}, {Bastian},
  {Biermann}, {Evans}, {Eyer}, {Jansen}, {Jordi}, {Klioner}, {Lammers},
  {Lindegren}, {Luri}, {Mignard}, {Milligan}, {Panem}, {Poinsignon},
  {Pourbaix}, {Randich}, {Sarri}, {Sartoretti}, {Siddiqui}, {Soubiran},
  {Valette}, {van Leeuwen}, {Walton}, {Aerts}, {Arenou}, {Cropper}, {Drimmel},
  {H{\o}g}, {Katz}, {Lattanzi}, {O'Mullane}, {Grebel}, {Holland}, {Huc},
  {Passot}, {Bramante}, {Cacciari}, {Casta{\~n}eda}, {Chaoul}, {Cheek}, {De
  Angeli}, {Fabricius}, {Guerra}, {Hern{\'a}ndez}, {Jean-Antoine-Piccolo},
  {Masana}, {Messineo}, {Mowlavi}, {Nienartowicz}, {Ord{\'o}{\~n}ez-Blanco},
  {Panuzzo}, {Portell}, {Richards}, {Riello}, {Seabroke}, {Tanga},
  {Th{\'e}venin}, {Torra}, {Els}, {Gracia-Abril}, {Comoretto},
  {Garcia-Reinaldos}, {Lock}, {Mercier}, {Altmann}, {Andrae}, {Astraatmadja},
  {Bellas-Velidis}, {Benson}, {Berthier}, {Blomme}, {Busso}, {Carry},
  {Cellino}, {Clementini}, {Cowell}, {Creevey}, {Cuypers}, {Davidson}, {De
  Ridder}, {de Torres}, {Delchambre}, {Dell'Oro}, {Ducourant}, {Fr{\'e}mat},
  {Garc{\'\i}a-Torres}, {Gosset}, {Halbwachs}, {Hambly}, {Harrison}, {Hauser},
  {Hestroffer}, {Hodgkin}, {Huckle}, {Hutton}, {Jasniewicz}, {Jordan},
  {Kontizas}, {Korn}, {Lanzafame}, {Manteiga}, {Moitinho}, {Muinonen},
  {Osinde}, {Pancino}, {Pauwels}, {Petit}, {Recio-Blanco}, {Robin}, {Sarro},
  {Siopis}, {Smith}, {Smith}, {Sozzetti}, {Thuillot}, {van Reeven}, {Viala},
  {Abbas}, {Abreu Aramburu}, {Accart}, {Aguado}, {Allan}, {Allasia},
  {Altavilla}, {{\'A}lvarez}, {Alves}, {Anderson}, {Andrei}, {Anglada Varela},
  {Antiche}, {Antoja}, {Ant{\'o}n}, {Arcay}, {Atzei}, {Ayache}, {Bach},
  {Baker}, {Balaguer-N{\'u}{\~n}ez}, {Barache}, {Barata}, {Barbier}, {Barblan},
  {Baroni}, {Barrado y Navascu{\'e}s}, {Barros}, {Barstow}, {Becciani},
  {Bellazzini}, {Bellei}, {Bello Garc{\'\i}a}, {Belokurov}, {Bendjoya},
  {Berihuete}, {Bianchi}, {Bienaym{\'e}}, {Billebaud}, {Blagorodnova},
  {Blanco-Cuaresma}, {Boch}, {Bombrun}, {Borrachero}, {Bouquillon}, {Bourda},
  {Bouy}, {Bragaglia}, {Breddels}, {Brouillet}, {Br{\"u}semeister},
  {Bucciarelli}, {Budnik}, {Burgess}, {Burgon}, {Burlacu}, {Busonero}, {Buzzi},
  {Caffau}, {Cambras}, {Campbell}, {Cancelliere}, {Cantat-Gaudin}, {Carlucci},
  {Carrasco}, {Castellani}, {Charlot}, {Charnas}, {Charvet}, {Chassat},
  {Chiavassa}, {Clotet}, {Cocozza}, {Collins}, {Collins}, {Costigan}, {Crifo},
  {Cross}, {Crosta}, {Crowley}, {Dafonte}, {Damerdji}, {Dapergolas}, {David},
  {David}, {De Cat}, {de Felice}, {de Laverny}, {De Luise}, {De March}, {de
  Martino}, {de Souza}, {Debosscher}, {del Pozo}, {Delbo}, {Delgado},
  {Delgado}, {di Marco}, {Di Matteo}, {Diakite}, {Distefano}, {Dolding}, {Dos
  Anjos}, {Drazinos}, {Dur{\'a}n}, {Dzigan}, {Ecale}, {Edvardsson}, {Enke},
  {Erdmann}, {Escolar}, {Espina}, {Evans}, {Eynard Bontemps}, {Fabre},
  {Fabrizio}, {Faigler}, {Falc{\~a}o}, {Farr{\`a}s Casas}, {Faye}, {Federici},
  {Fedorets}, {Fern{\'a}ndez-Hern{\'a}ndez}, {Fernique}, {Fienga}, {Figueras},
  {Filippi}, {Findeisen}, {Fonti}, {Fouesneau}, {Fraile}, {Fraser}, {Fuchs},
  {Furnell}, {Gai}, {Galleti}, {Galluccio}, {Garabato}, {Garc{\'\i}a-Sedano},
  {Gar{\'e}}, {Garofalo}, {Garralda}, {Gavras}, {Gerssen}, {Geyer}, {Gilmore},
  {Girona}, {Giuffrida}, {Gomes}, {Gonz{\'a}lez-Marcos},
  {Gonz{\'a}lez-N{\'u}{\~n}ez}, {Gonz{\'a}lez-Vidal}, {Granvik}, {Guerrier},
  {Guillout}, {Guiraud}, {G{\'u}rpide}, {Guti{\'e}rrez-S{\'a}nchez}, {Guy},
  {Haigron}, {Hatzidimitriou}, {Haywood}, {Heiter}, {Helmi}, {Hobbs},
  {Hofmann}, {Holl}, {Holland }, {Hunt}, {Hypki}, {Icardi}, {Irwin}, {Jevardat
  de Fombelle}, {Jofr{\'e}}, {Jonker}, {Jorissen}, {Julbe}, {Karampelas},
  {Kochoska}, {Kohley}, {Kolenberg}, {Kontizas}, {Koposov}, {Kordopatis},
  {Koubsky}, {Kowalczyk}, {Krone-Martins}, {Kudryashova}, {Kull}, {Bachchan},
  {Lacoste-Seris}, {Lanza}, {Lavigne}, {Le Poncin-Lafitte}, {Lebreton},
  {Lebzelter}, {Leccia}, {Leclerc}, {Lecoeur-Taibi}, {Lemaitre}, {Lenhardt},
  {Leroux}, {Liao}, {Licata}, {Lindstr{\o}m}, {Lister}, {Livanou}, {Lobel},
  {L{\"o}ffler}, {L{\'o}pez}, {Lopez-Lozano}, {Lorenz}, {Loureiro},
  {MacDonald}, {Magalh{\~a}es Fernandes}, {Managau}, {Mann}, {Mantelet},
  {Marchal}, {Marchant}, {Marconi}, {Marie}, {Marinoni}, {Marrese},
  {Marschalk{\'o}}, {Marshall}, {Mart{\'\i}n-Fleitas}, {Martino}, {Mary},
  {Matijevi{\v{c}}}, {Mazeh}, {McMillan}, {Messina}, {Mestre}, {Michalik},
  {Millar}, {Miranda}, {Molina}, {Molinaro}, {Molinaro}, {Moln{\'a}r},
  {Moniez}, {Montegriffo}, {Monteiro}, {Mor}, {Mora}, {Morbidelli}, {Morel},
  {Morgenthaler}, {Morley}, {Morris}, {Mulone}, {Muraveva}, {Musella},
  {Narbonne}, {Nelemans}, {Nicastro}, {Noval}, {Ord{\'e}novic},
  {Ordieres-Mer{\'e}}, {Osborne}, {Pagani}, {Pagano}, {Pailler}, {Palacin},
  {Palaversa}, {Parsons}, {Paulsen}, {Pecoraro}, {Pedrosa}, {Pentik{\"a}inen},
  {Pereira}, {Pichon}, {Piersimoni}, {Pineau}, {Plachy}, {Plum}, {Poujoulet},
  {Pr{\v{s}}a}, {Pulone}, {Ragaini}, {Rago}, {Rambaux}, {Ramos-Lerate},
  {Ranalli}, {Rauw}, {Read}, {Regibo}, {Renk}, {Reyl{\'e}}, {Ribeiro},
  {Rimoldini}, {Ripepi}, {Riva}, {Rixon}, {Roelens}, {Romero-G{\'o}mez},
  {Rowell}, {Royer}, {Rudolph}, {Ruiz-Dern}, {Sadowski}, {Sagrist{\`a}
  Sell{\'e}s}, {Sahlmann}, {Salgado}, {Salguero}, {Sarasso}, {Savietto},
  {Schnorhk}, {Schultheis}, {Sciacca}, {Segol}, {Segovia}, {Segransan},
  {Serpell}, {Shih}, {Smareglia}, {Smart}, {Smith}, {Solano}, {Solitro},
  {Sordo}, {Soria Nieto}, {Souchay}, {Spagna}, {Spoto}, {Stampa}, {Steele},
  {Steidelm{\"u}ller}, {Stephenson}, {Stoev}, {Suess}, {S{\"u}veges}, {Surdej},
  {Szabados}, {Szegedi-Elek}, {Tapiador}, {Taris}, {Tauran}, {Taylor},
  {Teixeira}, {Terrett}, {Tingley}, {Trager}, {Turon}, {Ulla}, {Utrilla},
  {Valentini}, {van Elteren}, {Van Hemelryck}, {van Leeuwen}, {Varadi},
  {Vecchiato}, {Veljanoski}, {Via}, {Vicente}, {Vogt}, {Voss}, {Votruba},
  {Voutsinas}, {Walmsley}, {Weiler}, {Weingrill}, {Werner}, {Wevers},
  {Whitehead}, {Wyrzykowski}, {Yoldas}, {{\v{Z}}erjal}, {Zucker}, {Zurbach},
  {Zwitter}, {Alecu}, {Allen}, {Allende Prieto}, {Amorim},
  {Anglada-Escud{\'e}}, {Arsenijevic}, {Azaz}, {Balm}, {Beck}, {Bernstein},
  {Bigot}, {Bijaoui}, {Blasco}, {Bonfigli}, {Bono}, {Boudreault}, {Bressan},
  {Brown}, {Brunet}, {Bunclark}, {Buonanno}, {Butkevich}, {Carret}, {Carrion},
  {Chemin}, {Ch{\'e}reau}, {Corcione}, {Darmigny}, {de Boer}, {de Teodoro}, {de
  Zeeuw}, {Delle Luche}, {Domingues}, {Dubath}, {Fodor}, {Fr{\'e}zouls},
  {Fries}, {Fustes}, {Fyfe}, {Gallardo}, {Gallegos}, {Gardiol}, {Gebran},
  {Gomboc}, {G{\'o}mez}, {Grux}, {Gueguen}, {Heyrovsky}, {Hoar}, {Iannicola},
  {Isasi Parache}, {Janotto}, {Joliet}, {Jonckheere}, {Keil}, {Kim},
  {Klagyivik}, {Klar}, {Knude}, {Kochukhov}, {Kolka}, {Kos}, {Kutka}, {Lainey},
  {LeBouquin}, {Liu}, {Loreggia}, {Makarov}, {Marseille}, {Martayan},
  {Martinez-Rubi}, {Massart}, {Meynadier}, {Mignot}, {Munari}, {Nguyen},
  {Nordlander}, {Ocvirk}, {O'Flaherty}, {Olias Sanz}, {Ortiz}, {Osorio},
  {Oszkiewicz}, {Ouzounis}, {Palmer}, {Park}, {Pasquato}, {Peltzer}, {Peralta},
  {P{\'e}turaud}, {Pieniluoma}, {Pigozzi}, {Poels}, {Prat}, {Prod'homme},
  {Raison}, {Rebordao}, {Risquez}, {Rocca-Volmerange}, {Rosen}, {Ruiz-Fuertes},
  {Russo}, {Sembay}, {Serraller Vizcaino}, {Short}, {Siebert}, {Silva},
  {Sinachopoulos}, {Slezak}, {Soffel}, {Sosnowska}, {Strai{\v{z}}ys}, {ter
  Linden}, {Terrell}, {Theil}, {Tiede}, {Troisi}, {Tsalmantza}, {Tur},
  {Vaccari}, {Vachier}, {Valles}, {Van Hamme}, {Veltz}, {Virtanen}, {Wallut},
  {Wichmann}, {Wilkinson}, {Ziaeepour}, \&
  {Zschocke}}]{Gaia_Collaboration_2016}
{Gaia Collaboration (Prusti, T., et~al.)} 2016,
  \aap, 595, A1

\bibitem[{{Galv{\'a}n-Madrid} {et~al.}(2015){Galv{\'a}n-Madrid},
  {Rodr{\'{\i}}guez}, {Liu}, {Costigan}, {Palau}, {Zapata}, \&
  {Loinard}}]{galvan-madrid15}
{Galv{\'a}n-Madrid}, R., {Rodr{\'{\i}}guez}, L.~F., {Liu}, H.~B., {et~al.}
  2015, \apjl, 806, L32

\bibitem[{{Getman} {et~al.}(2017){Getman}, {Broos}, {Kuhn}, {Feigelson},
  {Richert}, {Ota}, {Bate}, \& {Garmire}}]{getman17}
{Getman}, K.~V., {Broos}, P.~S., {Kuhn}, M.~A., {et~al.} 2017, \apjs, 229, 28

\bibitem[{{Getman} {et~al.}(2008){Getman}, {Feigelson}, {Broos}, {Micela}, \&
  {Garmire}}]{getman08}
{Getman}, K.~V., {Feigelson}, E.~D., {Broos}, P.~S., {Micela}, G., \&
  {Garmire}, G.~P. 2008, \apj, 688, 418

\bibitem[{{Getman} {et~al.}(2010){Getman}, {Feigelson}, {Broos}, {Townsley}, \&
  {Garmire}}]{getman10}
{Getman}, K.~V., {Feigelson}, E.~D., {Broos}, P.~S., {Townsley}, L.~K., \&
  {Garmire}, G.~P. 2010, \apj, 708, 1760

\bibitem[{{Getman} {et~al.}(2007){Getman}, {Feigelson}, {Garmire}, {Broos}, \&
  {Wang}}]{getman07}
{Getman}, K.~V., {Feigelson}, E.~D., {Garmire}, G., {Broos}, P., \& {Wang}, J.
  2007, \apj, 654, 316

\bibitem[{{Giardino} {et~al.}(2007){Giardino}, {Favata}, {Micela}, {Sciortino},
  \& {Winston}}]{giardino07b}
{Giardino}, G., {Favata}, F., {Micela}, G., {Sciortino}, S., \& {Winston}, E.
  2007, \aap, 463, 275

\bibitem[{{Goodman} \& {Weare}(2010)}]{goodman10}
{Goodman}, J. \& {Weare}, J. 2010, Commun. Appl. Math. Comput. Sci., 5, 65

\bibitem[{{Gounelle} {et~al.}(2013){Gounelle}, {Chaussidon}, \&
  {Rollion-Bard}}]{gounelle13}
{Gounelle}, M., {Chaussidon}, M., \& {Rollion-Bard}, C. 2013, \apjl, 763, L33

\bibitem[{{Groppi} {et~al.}(2007){Groppi}, {Hunter}, {Blundell}, \&
  {Sandell}}]{groppi07}
{Groppi}, C.~E., {Hunter}, T.~R., {Blundell}, R., \& {Sandell}, G. 2007, \apj,
  670, 489

\bibitem[{{G{\"u}del}(2002)}]{guedel02}
{G{\"u}del}, M. 2002, \araa, 40, 217

\bibitem[{{G{\"u}del} {et~al.}(2007){G{\"u}del}, {Briggs}, {Arzner}, {Audard},
  {Bouvier}, {Feigelson}, {Franciosini}, {Glauser}, {Grosso}, {Micela},
  {Monin}, {Montmerle}, {Padgett}, {Palla}, {Pillitteri}, {Rebull}, {Scelsi},
  {Silva}, {Skinner}, {Stelzer}, \& {Telleschi}}]{guedel07}
{G{\"u}del}, M., {Briggs}, K.~R., {Arzner}, K., {et~al.} 2007, \aap, 468, 353

\bibitem[{{G{\"u}del} \& {Naz{\'e}}(2009)}]{guedel09}
{G{\"u}del}, M. \& {Naz{\'e}}, Y. 2009, \aapr, 17, 309

\bibitem[{{Hamaguchi} {et~al.}(2005){Hamaguchi}, {Corcoran}, {Petre}, {White},
  {Stelzer}, {Nedachi}, {Kobayashi}, \& {Tokunaga}}]{hamaguchi05}
{Hamaguchi}, K., {Corcoran}, M.~F., {Petre}, R., {et~al.} 2005, \apj, 623, 291

\bibitem[{{Hamaguchi} {et~al.}(2010){Hamaguchi}, {Grosso}, {Kastner},
  {Weintraub}, \& {Richmond}}]{hamaguchi10}
{Hamaguchi}, K., {Grosso}, N., {Kastner}, J.~H., {Weintraub}, D.~A., \&
  {Richmond}, M. 2010, \apjl, 714, L16

\bibitem[{{Hartmann} {et~al.}(1994){Hartmann}, {Hewett}, \&
  {Calvet}}]{hartmann94}
{Hartmann}, L., {Hewett}, R., \& {Calvet}, N. 1994, \apj, 426, 669

\bibitem[{{Hogg} \& {Foreman-Mackey}(2018)}]{hogg18}
{Hogg}, D.~W. \& {Foreman-Mackey}, D. 2018, \apjs, 236, 11

\bibitem[{{Kamezaki} {et~al.}(2014){Kamezaki}, {Imura}, {Omodaka}, {Handa},
  {Tsuboi}, {Nagayama}, {Hirota}, {Sunada}, {Kobayashi}, {Chibueze}, {Kawai},
  \& {Nakano}}]{kamezaki14}
{Kamezaki}, T., {Imura}, K., {Omodaka}, T., {et~al.} 2014, \apjs, 211, 18

\bibitem[{{Kastner} {et~al.}(2005){Kastner}, {Franz}, {Grosso}, {Bally},
  {McCaughrean}, {Getman}, {Feigelson}, \& {Schulz}}]{kastner05}
{Kastner}, J.~H., {Franz}, G., {Grosso}, N., {et~al.} 2005, \apjs, 160, 511

\bibitem[{{Krabbendam} {et~al.}(2004){Krabbendam}, {Heathcote}, {Schumacher},
  {Schwarz}, {Sebring}, \& {Warner}}]{krabbendam04}
{Krabbendam}, V.~L., {Heathcote}, S., {Schumacher}, G., {et~al.} 2004, 5489,
  615

\bibitem[{{Loh} {et~al.}(2012){Loh}, {Biel}, {Davis}, {Laporte}, {Loh}, \&
  {Verhanovitz}}]{loh12}
{Loh}, E.~D., {Biel}, J.~D., {Davis}, M.~W., {et~al.} 2012, \pasp, 124, 343

\bibitem[{{Mainzer} {et~al.}(2014){Mainzer}, {Bauer}, {Cutri}, {Grav},
  {Masiero}, {Beck}, {Clarkson}, {Conrow}, {Dailey}, {Eisenhardt}, {Fabinsky},
  {Fajardo-Acosta}, {Fowler}, {Gelino}, {Grillmair}, {Heinrichsen}, {Kendall},
  {Kirkpatrick}, {Liu}, {Masci}, {McCallon}, {Nugent}, {Papin}, {Rice},
  {Royer}, {Ryan}, {Sevilla}, {Sonnett}, {Stevenson}, {Thompson}, {Wheelock},
  {Wiemer}, {Wittman}, {Wright}, \& {Yan}}]{mainzer14}
{Mainzer}, A., {Bauer}, J., {Cutri}, R.~M., {et~al.} 2014, \apj, 792, 30

\bibitem[{{Marrese} {et~al.}(2019){Marrese}, {Marinoni}, {Fabrizio}, \&
  {Altavilla}}]{marrese19}
{Marrese}, P.~M., {Marinoni}, S., {Fabrizio}, M., \& {Altavilla}, G. 2019,
  \aap, 621, A144

\bibitem[{{Megeath} {et~al.}(2012){Megeath}, {Gutermuth}, {Muzerolle},
  {Kryukova}, {Flaherty}, {Hora}, {Allen}, {Hartmann}, {Myers}, {Pipher},
  {Stauffer}, {Young}, \& {Fazio}}]{megeath12}
{Megeath}, S.~T., {Gutermuth}, R., {Muzerolle}, J., {et~al.} 2012, \aj, 144,
  192

\bibitem[{{Navarete} {et~al.}(2015){Navarete}, {Damineli}, {Barbosa}, \&
  {Blum}}]{navarete15}
{Navarete}, F., {Damineli}, A., {Barbosa}, C.~L., \& {Blum}, R.~D. 2015,
  \mnras, 450, 4364

\bibitem[{{Ormel} {et~al.}(2011){Ormel}, {Min}, {Tielens}, {Dominik}, \&
  {Paszun}}]{ormel11}
{Ormel}, C.~W., {Min}, M., {Tielens}, A.~G.~G.~M., {Dominik}, C., \& {Paszun},
  D. 2011, \aap, 532, A43

\bibitem[{{Palla} \& {Stahler}(1992)}]{palla92}
{Palla}, F. \& {Stahler}, S.~W. 1992, \apj, 392, 667

\bibitem[{{Palla} \& {Stahler}(1999)}]{palla99}
{Palla}, F. \& {Stahler}, S.~W. 1999, \apj, 525, 772

\bibitem[{{Pillitteri} {et~al.}(2019){Pillitteri}, {Sciortino}, {Reale},
  {Micela}, {Argiroffi}, {Flaccomio}, \& {Stelzer}}]{pillitteri19}
{Pillitteri}, I., {Sciortino}, S., {Reale}, F., {et~al.} 2019, \aap, 623, A67

\bibitem[{{Reipurth} \& {Bally}(2001)}]{reipurth01}
{Reipurth}, B. \& {Bally}, J. 2001, \araa, 39, 403

\bibitem[{{Rodr{\'\i}guez} {et~al.}(2017){Rodr{\'\i}guez}, {Zapata}, \&
  {Palau}}]{rodriguez17b}
{Rodr{\'\i}guez}, L.~F., {Zapata}, L.~A., \& {Palau}, A. 2017, \aj, 153, 209

\bibitem[{{Safron} {et~al.}(2015){Safron}, {Fischer}, {Megeath}, {Furlan},
  {Stutz}, {Stanke}, {Billot}, {Rebull}, {Tobin}, {Ali}, {Allen}, {Booker},
  {Watson}, \& {Wilson}}]{safron15}
{Safron}, E.~J., {Fischer}, W.~J., {Megeath}, S.~T., {et~al.} 2015, \apjl, 800,
  L5

\bibitem[{{Saraceno} {et~al.}(1996){Saraceno}, {Andr\'e}, {Ceccarelli},
  {Griffin}, \& {Molinari}}]{saraceno96}
{Saraceno}, P., {Andr\'e}, P., {Ceccarelli}, C., {Griffin}, M., \& {Molinari},
  S. 1996, \aap, 309, 827

\bibitem[{{Schirmer}(2013)}]{schirmer13}
{Schirmer}, M. 2013, \apjs, 209, 21

\bibitem[{{Shang} {et~al.}(2004){Shang}, {Lizano}, {Glassgold}, \&
  {Shu}}]{shang04}
{Shang}, H., {Lizano}, S., {Glassgold}, A., \& {Shu}, F. 2004, \apjl, 612, L69

\bibitem[{{Smith} {et~al.}(2001){Smith}, {Brickhouse}, {Liedahl}, \&
  {Raymond}}]{smith01}
{Smith}, R.~K., {Brickhouse}, N.~S., {Liedahl}, D.~A., \& {Raymond}, J.~C.
  2001, \apjl, 556, L91

\bibitem[{{Sossi} {et~al.}(2017){Sossi}, {Moynier}, {Chaussidon}, {Villeneuve},
  {Kato}, \& {Gounelle}}]{sossi17}
{Sossi}, P.~A., {Moynier}, F., {Chaussidon}, M., {et~al.} 2017, Nature
  Astronomy, 1, 0055

\bibitem[{{Stahler}(1988)}]{stahler88}
{Stahler}, S.~W. 1988, \apj, 332, 804

\bibitem[{{Stanke} {et~al.}(2002){Stanke}, {McCaughrean}, \&
  {Zinnecker}}]{stanke02}
{Stanke}, T., {McCaughrean}, M.~J., \& {Zinnecker}, H. 2002, \aap, 392, 239

\bibitem[{{Taylor}(2019)}]{taylor19}
{Taylor}, M.~B. 2019, in Astronomical Data Analysis
  Software and Systems XXVII, ed. P.~J. {Teuben}, M.~W. {Pound}, B.~A.
  {Thomas}, \& E.~M. {Warner},
  ASP Conf.\ Ser., 523, 43

\bibitem[{{Tsujimoto} {et~al.}(2005){Tsujimoto}, {Feigelson}, {Grosso},
  {Micela}, {Tsuboi}, {Favata}, {Shang}, \& {Kastner}}]{tsujimoto05}
{Tsujimoto}, M., {Feigelson}, E.~D., {Grosso}, N., {et~al.} 2005, \apjs, 160,
  503

\bibitem[{{Tsujimoto} {et~al.}(2002){Tsujimoto}, {Koyama}, {Tsuboi}, {Goto}, \&
  {Kobayashi}}]{tsujimoto02}
{Tsujimoto}, M., {Koyama}, K., {Tsuboi}, Y., {Goto}, M., \& {Kobayashi}, N.
  2002, \apj, 566, 974

\bibitem[{{Weisskopf} {et~al.}(2002){Weisskopf}, {Brinkman}, {Canizares},
  {Garmire}, {Murray}, \& {Van Speybroeck}}]{weisskopf02}
{Weisskopf}, M.~C., {Brinkman}, B., {Canizares}, C., {et~al.} 2002, \pasp, 114,
  1

\bibitem[{{Whitney} {et~al.}(2003){Whitney}, {Wood}, {Bjorkman}, \&
  {Wolff}}]{whitney03}
{Whitney}, B.~A., {Wood}, K., {Bjorkman}, J.~E., \& {Wolff}, M.~J. 2003, \apj,
  591, 1049

\bibitem[{{Wilms} {et~al.}(2000){Wilms}, {Allen}, \& {McCray}}]{wilms00}
{Wilms}, J., {Allen}, A., \& {McCray}, R. 2000, \apj, 542, 914

\bibitem[{{Wright} {et~al.}(2010){Wright}, {Eisenhardt}, {Mainzer}, {Ressler},
  {Cutri}, {Jarrett}, {Kirkpatrick}, {Padgett}, {McMillan}, {Skrutskie},
  {Stanford}, {Cohen}, {Walker}, {Mather}, {Leisawitz}, {Gautier}, {McLean},
  {Benford}, {Lonsdale}, {Blain}, {Mendez}, {Irace}, {Duval}, {Liu}, {Royer},
  {Heinrichsen}, {Howard}, {Shannon}, {Kendall}, {Walsh}, {Larsen}, {Cardon},
  {Schick}, {Schwalm}, {Abid}, {Fabinsky}, {Naes}, \& {Tsai}}]{wright10b}
{Wright}, E.~L., {Eisenhardt}, P.~R.~M., {Mainzer}, A.~K., {et~al.} 2010, \aj,
  140, 1868

\end{thebibliography}
\end{document}